\documentclass[letterpaper, 10pt, conference]{ieeeconf}
\usepackage{enumerate}
\usepackage{bm}

\usepackage{ifthen}
\usepackage{lineno}
\usepackage{amssymb}
\usepackage{amsmath}
\usepackage{adjustbox}
\usepackage{epsfig}
\usepackage{graphicx}
\usepackage{graphics}
\usepackage{float}
\usepackage{bbm}
\usepackage{color}
\usepackage{soul}

\renewcommand{\P}{{\mathcal P}}
\newcommand{\x}{{\bf x}}
\newcommand{\xs}{ x}

\newcommand{\Proofs}[2]{#1}

\newcommand{\eop} {\hfill{$\blacksquare$}}
\newcommand{\Cmnt}[1] {}
\newtheorem{theorem}{Theorem}

\newtheorem{lemma}{Lemma}

\definecolor{darkred}{rgb}{1, 0.1, 0.3}
\definecolor{darkblue}{rgb}{0.1, 0.1, 1}
\definecolor{darkgreen}{rgb}{0,0.6,0.5}

\newcommand {\mm}[1] {\ifmmode{#1}\else{\mbox{\(#1\)}}\fi}

\newcommand{\AsymCmnt}[1]{}   % Just for commenting out Asymmetric player results as of now..

\newcommand{\R}{\mathbb{R}}

\newcommand{\Exp}[1]{\mathbb{E}\left[#1\right]} %Expectation

\newcommand{\hide}[1]{}
\DeclareMathOperator*{\argmax}{arg\,max}

\newboolean{showcomments}
\setboolean{showcomments}{true}
\newcommand{\jk}[1]{\ifthenelse{\boolean{showcomments}} {\textcolor{red}{(JK says: #1)}} {} }
\renewcommand{\ss}[1]{\ifthenelse{\boolean{showcomments}} {\textcolor{red}{(SS says: #1)}} {} }
\newcommand{\vk}[1]{\ifthenelse{\boolean{showcomments}} {\textcolor{red}{(VK says: #1)}} {} }

\makeatletter
\let\NAT@parse\undefined
\makeatother
\usepackage{hyperref}

\begin{document}

\title{Coalition Formation in Constant Sum Queueing Games} 
\author{
$\text{ Shiksha Singhal}^1$, $\text{Veeraruna Kavitha}^1$ and $\text{Jayakrishnan Nair}^2$ \\
$\text{ IEOR}^1$, $\text{EE}^2$, Indian Institute of Technology Bombay, India
}

% \institute{Stockholm University, Stockholm, Sweden\\\email{elena.touli@math.su.se}\andthe Ohio State University Columbus, Ohio, U.S.A.\\\email{yusu@cse.ohio-state.edu}\\}\\

%\authorrunning{Mokhov, Sutcliffe and Voronkov}

% \title{FPT-Algorithms for computing Gromov-Hausdorff and interleaving distances between trees}
% \author{Elena Farahbkhsh Touli} \and \author{y}
%\date{}

%\begin{document}
\maketitle
%\linenumbers
\setcounter{page}{1}

\begin{abstract}
  We analyse a coalition formation game between strategic service
  providers of a congestible service. The key novelty of our
  formulation is that it is a constant sum game, i.e., the total
  payoff across all service providers (or coalitions of providers) is
  fixed, and dictated by the total size of the market. The game thus
  captures the tension between resource pooling (to benefit from the
  resulting statistical economies of scale) and competition between
  coalitions over market share. In a departure from the prior
  literature on resource pooling for congestible services, we show
  that the grand coalition is in general not stable, once we allow for
  competition over market share. Instead, the stable configurations
  are duopolies, where the dominant coalition exploits its economies
  of scale to corner a disproportionate market share. We analyse the
  stable duopolies that emerge from this interaction, and also study a
  dynamic variant of this game.
\end{abstract}

\section{Introduction}

%\section{Introduction}
Resource sharing is an efficient way of reducing congestion and
uncertainty in service industries. It
refers to an arrangement where service resources are pooled and used
jointly by a group (a.k.a., coalition) of providers, instead of each
provider operating alone using its own resources. Naturally, such a
coalition would be sustainable only if the participating providers
obtain higher payoffs than they would have obtained otherwise. The key
driver of coalition formation in congestion prone service systems is
the statistical economies of scale that emerge from the pooling of
service resources---this allows the coalition to offer a better
quality of service to its customers, and/or to attract more customers
to its service.
%One of the important advantage of resource sharing is the reduction
%in number of customers who are denied service (or blocked) due to the
%unavailability of resources, leading to increased customer
%satisfaction. Another benefit is exploiting economies of scale that
%arise due to resource sharing.
\hide{
There are numerous applications where independent service providers
consider resource pooling to improve their payoffs. Some such
applications include sharing of operating rooms between various
departments in a hospital, sharing of spectrum between cellular
operators, and sharing of ambulances or fire fighting machinery
between neighboring districts.
}

Not surprisingly, there is a considerable literature that analyses
resource pooling between independent providers of congestible services
via a cooperative game theoretic approach. In these papers, each
provider is modeled as a queueing system, with its own dedicated
customer base, that generates service requests according to a certain
arrival process. The payoff of each service provider is in turn
determined by the quality of service it is able to provide to its
(dedicated) customer base. In such a setting, the statistical
economies of scale from resource pooling typically drives the service
providers to pool all their servers together to form a \emph{grand
coalition}, which generates the greatest aggregate payoff across all
coalitional arrangements. Naturally, the resulting aggregate payoff
must be divided between the providers in a \emph{stable} manner, i.e.,
in such a way that no subset of providers has an incentive to `break
away' from the grand coalition. Such stable payoff allocations have
been demonstrated in a wide range of settings, including
single/multiple server environments, and loss/queue-based environments
(see \cite{karsten,karsten_loss}  and the references therein).

To summarize, the literature on coalition formation between providers
of congestible services suggests that a stable grand coalition would
emerge from the strategic interaction. However, a crucial aspect the
preceding literature fails to capture is \emph{user churn}. That is,
customers can switch service providers, if offered superior service
quality elsewhere. This aspect introduces \emph{competition} between
the service providers (or coalitions of service providers) over market
share. To the best of our knowledge, the interplay between resource
pooling among service providers (and the associated economies of
scale) with the competition between them, in the context of
congestible services, has not been explored in the literature. This
paper seeks to fill this gap.

Specifically, we analyse a coalition formation game between a
collection of service providers, each of which is modelled as an
Erlang-B loss system. A key aspect of our model is that the total
market size (captured via the aggregate arrival rate of customer
requests) is fixed exogenously, making the game \emph{constant sum},
i.e., the total payoff across all providers (or coalitions of
providers) is fixed. This constant sum aspect, as we show,
dramatically alters the outcome of the strategic interaction between
providers. In particular, we show that (except in a very specific
corner case), \emph{the grand coalition is not} stable. Instead, the
predominant stable configurations are \emph{duopolies}, with the
larger coalition exploiting economies of scale to corner a
disproportionate portion of the market share. Our work also highlights
several subtleties relating to different natural notions of stability
in this context, the way the payoff of each coalition is divided
between its members, and the degree of congestion in the system.

Our contributions are as follows.

\noindent 1. We formally define a constant sum coalition formation
game between strategic service providers of a congestible
service. This model is the first, to the best of our knowledge, to
capture the interplay between resource pooling and
competition. Crucially, this is a \emph{partition form} game, since
the payoff of each coalition depends on not just the members of that
coalition, but also on the coalitional arrangements outside the
coalition.

\noindent 2. We introduce three natural definitions of a \emph{stable
configuration} for this game, where a configuration specifies a
partition of the set of providers (into coalitions), and also the
allocation of the total payoff of each coalition among its
members. The three notions of stability we consider differ with
respect to the range of deviations or movements that are blocked (or
disincentivised), and also the precision with which the coalitions
that seek to `break' from the prevailing configuration can estimate
the benefit from doing so.

\noindent 3. We analyse the class of stable configurations that emerge
under each notion of stability. Interestingly, we are able to show
that under configurations involving three or more coalitions are not
stable under any notion of stability. Intuitively, this is because the
economies of scale that incentivise certain mergers to take place
between coalitions. Moreover, except for a corner case, we show that
the grand coalition also cannot be part of a stable
configuration. This means the dominant equilibria for this system are
duopolies.

\noindent 4. We further analyse the stable configurations that involve
duopolies under each stability notion. Interestingly, the payoff
allocations supported by stable configurations differ across the
different stability notions. We also explore the impact of the overall
congestion level on the stable duopolies, by analysing light and heavy
traffic regimes.

\noindent 5. Finally, we study a dynamic variant of the coalition
formation game, and analyse the conditions for the (random) dynamics
to `settle' to a stable configuration in a finite number of moves.

\section{Model and Preliminaries}
\label{sec_model}

In this section, we describe our system model for coalition formation
between strategic service providers, characterize the behavior of the
customer base in response to coalition formation between service
providers, and introduce some background on the notion of a
\emph{stable configuration}.

\subsection{System model}
Consider a system with a set $\mathcal{N} = \{1,\cdots,n\}$ of
independent service providers (a.k.a., agents), with provider~$i$
having $N_i$ servers. Without loss of generality, we assume
$N_i \ge N_{i+1}$ for $1 \leq i \leq n-1.$ All servers are identical,
and assumed to have a unit speed, without loss of generality. The
providers serve a customer base that generates service requests as per
a Poisson process of rate~$\Lambda.$ Jobs sizes (a.k.a., service
requirements) are i.i.d., with~$J$ denoting a generic job size, and
$\Exp{J} = 1/\mu.$
% {\color{red} Erlang-B formula holds for any distribution with means
% $1/\mu$}. The service times are exponentially distributed with rate
% $\mu$.

Service providers are strategic, and can form
coalitions with other service providers to enhance their
rewards. Formally, such coalition formation between the service
providers induces a partition~$\P = \{C_1,C_2,\cdots,C_k\}$ of
$\mathcal{N},$ where
$$\cup_{i = 1}^kC_i = \mathcal{N},\quad C_i \cap C_j = \emptyset \ \forall\ i
\neq j.$$ We refer to such a partition with $k$ coalitions as a
$k$-partition. (Naturally, the baseline scenario where each service
provider operates independently corresponds to an $n$-partition.)

In response to a partition~$\P$ induced by coalition formation between
service providers, the arrival process of customer requests gets split
across the~$k$ coalitions in $\P$, with the arrival process seen by
coalition~$C$ being a Poisson process of rate $\lambda^{\P}_{C},$
where $\sum_{C \in \P} \lambda^{\P}_{C} = \Lambda.$ (We characterize
the split $(\lambda^{\P}_{C},\ C \in \P)$ as a Wardrop equilibrium;
details below.) Each coalition~$C$ operates as an
$M$/$M$/$N_{C}$/$N_{C}$ (Erlang-B) loss system, with
$N_{C} = \sum_{j \in C} N_j$ parallel servers, and arrival
rate~$\lambda^{\P}_{C}.$ This means jobs arriving into coalition~$C$
that find a free server upon arrival begin service immediately, while
those that arrive when all $N_{C}$ servers are busy get dropped
(lost). Given the well known insensitivity property of the Erlang-B
system, the steady state blocking probability associated with
coalition~$C$ (the long run fraction of jobs arriving into
coalition~$C$ that get dropped), denoted $B_{C}^{\P},$ is given by the
Erlang-B formula (\cite{formulas}):
% The customers divide themselves among
% the independent operating units of the system. Each provider operates
% as an $M/M/N/N$ queue, i.e., Erlang loss system, when it is alone.
% The providers are also seeking collaboration possibilities with some
% of the others.  If some of them operate together, for example if
% providers $i$ and $j$ form coalition $C = \{i, j\}$ then we will have
% an $M/M/N_C/N_C$ queue with $N_C = \sum_{l \in C} N_l$. In this case,
% one will have joint arrivals to this combined unit.  Using the well
% known Erlang-B formula, if the providers in $C$ operate together
% and %if they manage to
% attract customers at rate $\lambda_C$, the steady-state blocking
% probability $B_C$ of this unit is \vspace{-1mm}

\vspace{-5mm}
{\small \begin{align}
\label{Eqn_PB}
&B_{C}^{\P} = B(N_{C},a^{\P}_{C}), \text{ where } a^{\P}_{C} := \frac{\lambda^{\P}_{C}}{\mu},\\ 
%\displaystyle
& B(N,a) = \frac{ \frac{a^{N}}{N!}   }{ \sum_{j=0}^{N} \frac{a^{j}}{j!} }. \nonumber
\end{align}}

\hide{
Let ${\cal P} = \{ C_1, \cdots, C_k\}$ represent any partition of
$\mathcal{N}$, {\color{blue}referred to as $k$-partition}, which
describes the subsets of providers that are operating together.
Observe here any partition satisfies
\begin{equation}
\cup_{l=1}^k C_l = \mathcal{N} \text{ and } C_l \cap C_j = \emptyset \, \, \forall \, l \neq j.
\label{partition_define}
\end{equation}
Given a partition as above, it implies all providers in any $C$ are
operating together and have joint arrivals.  It is obvious that the
fraction of customers that choose (say) unit $C$ also depends upon the
partition of the rest of the players ${\cal N} \backslash C$, i.e.,
upon ${\cal P}$.  As is usually considered in these kind of scenarios,
we assume that customers split into the $k$ units of ${\cal P}$
according to the well-known equilibrium concept, Wardrop Equilibrium.
}%end hide

\subsection{User behavior: Wardrop equilibrium}

Next, we define the behavior of the customer base in response to
coalition formation across service providers, via the split
$(\lambda^{\P}_{C},\ C \in \P)$ of the aggregate arrival process of
service requests across coalitions. This split is characterized as a
Wardrop equilibrium (\cite{WE}).

\Cmnt{
The Wardrop equilibrium (WE)  is extensively used to predict
traffic patterns in transportation networks that are subject to
congestion. Its defining principle is that the user traffic splits
itself among the available routes in such a way that the journey times
along all used routes are equal, and less than those that would be
experienced on any unused route. Thus, the WE may be interpreted as a
Nash equilibrium between infinitesimal, non-atomic users. \jk{This
  para can be removed later if we face space constraints.} }
% This is often referred to as ``User Equilibrium".  The second
% principle is concerned with {\color{red}``System Optimality"} and
% states that at equilibrium, the average journey time is at a
% minimum.
%% JK: The second principle is an *alternative* model for traffic
%% proposed by Wardrop, distinct from the WE.

% {\color{red} Define WE in our context.}
In the context of our model, we define the WE split of the arrival
process of service requests across coalitions, such that the steady
state blocking probability associated with each coalition is equal.
% We consider a similar model where customers can choose any one of
% the $k$ units in $\mathcal{P}$.  With blocking probability as the
% cost of each unit (each coalition of the partition), customers split
% themselves among the $k$ units in such a way the the blocking
% probability of each used unit is equal.
Note that since the blocking probability associated with an `unused'
coalition would be zero, it follows that all coalitions would see a
strictly positive arrival rate. Thus, the WE (if it exists) is
characterized by a vector of arrival rates
$(\lambda^{\P}_{C},\ C \in \P)$ satisfying,
% {\color{red}If customers do not choose some of the units in
% $\mathcal{P}$, those units will have $0$ blocking probability which
% is smaller than that of the used ones and hence all units are
% used. } Also, customers obtain smaller blocking probabilities in
% this case. Hence, customers split themselves amongst all the $k$
% units in partition
% $\mathcal{P}$. % one can define WE for our problem for any partition ${\cal P}$, as shown below:
\begin{equation}
  \label{Eqn_WE_properties}
  B^{\P}_C = B\left(N_C,\frac{\lambda_C^\P}{\mu} \right) = B^*  \ \forall\ C \in \P, \quad 
 \sum_{C \in  \P } \lambda_C^\P  = \Lambda ,
\end{equation}
where $B^*$ is the common steady state blocking probability for each
coalition. For any given partition~$\P,$ the following theorem
establishes the existence and uniqueness of the WE, along with some
useful properties.
\begin{theorem}
  \label{Thm_WE}
 \textit{ Given any partition~$\P$ between the service providers, there is a
  unique Wardrop equilibrium $(\lambda^{\P}_{C},\ C \in \P),$ where
  $\lambda^{\P}_{C} > 0$ for all $C \in \P,$ that satisfies
  \eqref{Eqn_WE_properties}. Additionally, the following properties hold:\\
  $(i)$ For each~$C \in \P,$ $\lambda_{C}^\P$ is a strictly increasing
  function of the total arrival rate $\Lambda.$\\
  $(ii)$  If the partition $\P'$ is formed by merging two coalitions
  $C_i$ and $C_j$ in partition $\P$ where
  $C_i \cup C_j \neq \mathcal{N}$ (with all other coalitions in~$\P$
  remaining intact),
  $$\lambda^{\P'}_{C_i \cup C_j} > \lambda^{\P}_{C_i} +
  \lambda^{\P}_{C_j}.$$
   \Cmnt{b) Similarly, if $\P'  $ is formed by  split of one of the coalition $C_j$ into $S$ and $S'$ (with all other coalitions in~$\P$
  remaining intact) then 
$$
  \lambda_{S}^{\P'} +   \lambda_{S'}^{\P'} < \lambda_{C_j}^{\P}.
$$ }
  $(iii)$ If $\P = \{C_1,C_2\},$ with $N_{C_1} > N_{C_2},$
  then
  $$\frac{\lambda^{\P}_{C_1}}{N_{C_1}} > \frac{\Lambda}{N} >
  \frac{\lambda^{\P}_{C_2}}{N_{C_2}}, \text{ where } N = \sum_{i \in
    \mathcal{N}} N_i.$$}
 
 \end{theorem}
 \textbf{Proof: }See Appendix B. \eop %\jk{Is $(ii)b$ not a simply a
   %corollary of $(ii)a$?}
 
 Aside from asserting the uniqueness and strict positivity of the
 Wardrop split, Theorem~\ref{Thm_WE} also states that equilibrium
 arrival rate of each coalition is an increasing function of the
 aggregate arrival rate~$\Lambda;$ see Statement~$(i).$ Additionally,
 Statement~$(ii)$ demonstrates the statistical economies of scale due
 to a merger between coalitions: the merged entity is able to attract
 an arrival rate that exceeds the sum of the arrival rates seen by the
 two coalitions pre-merger. Finally, Statement~$(iii)$ provides
 another illustration of statistical economies of scale for the
 special case of a 2-partition---the larger coalition enjoys a higher
 utilization per server than the smaller one. %The proof of
 %Theorem~\ref{Thm_WE} can be found in Appendix~\ref{??}.

% \textbf{Proof: }See Appendix B. \eop

 \Cmnt{ We use WE to find the individual arrival rates
   $\lambda_C^\mathcal{P}$ such that the blocking probability of each
   coalition $C$ in partition $\mathcal{P}$ is same and hence follow
   its first principle.
   
   \subsubsection*{Feasible Allocations at WE}
   The allocation at WE is said to be feasible if the blocking
   probability of each coalition in the partition is same and the sum
   of arrival rates to any coalition $C$ is equal to the total arrival
   rate, i.e.,
%\vspace{-2mm}

   where $N_C = \sum_{i \in C} N_i$.  We now state some important
   properties of the WE.  The following lemma characterises the
   \textit{monotonicity property} of the WE:
   \begin{lemma}
     \textit{Assuming the existence of WE, the individual arrival
       rates $\lambda_i \, \forall \, i \in \mathcal{N}$ at WE is an
       increasing function of the total arrival rate $\Lambda$. }
     \label{WE_2waysplit_monotonic}
   \end{lemma} 
   \noindent\textbf{Proof:}
   See Appendix. \eop
   
   The next result proves the \textit{existence} of WE. Moreover, we have the \textit{uniqueness}.
   
   \begin{lemma}
     \textit{Under the proposed queuing model, WE exists and is also unique.}
     \label{exist_and_unique}
   \end{lemma}
   \textbf{Proof: }See Appendix B. \eop
 }
 
 \subsection{Coalition formation game: Preliminaries}
 
 Having defined the behavior of the user base, we now provide some
 preliminary details on the coalition formation game between the
 service providers.

 Recall, that each service provider is strategic, and only enters into
 a coalition if doing so is beneficial.
 % An equilibrium of this game will be characterized by a \emph{stable
 % configuration}, which is specified by a partition of the service
 % providers, and an allocation of payoff to each service
 % provider. \jk{Abberviate by SP?  Or use `player' instead?} 
 Given a partition ${\cal P}$ that describes the coalitions formed by
 the service providers, we define the value or payoff of each
 coalition $C \in \P$ to be $\beta \lambda_C^{\P},$ where $\beta > 0.$
 This is of course natural when the coalition derives a certain
 revenue per served job. The same model is also applicable if
 $\lambda_C^{\P}$ is interpreted as being proportional to the number of
 subscribers of coalition~$C,$ with each subscriber paying a recurring
 subscription fee. Without loss of generality, we set~$\beta = 1.$
 
 % each unit/coalition attracts customers at $\lambda_C^P$.  Thus the
 % value of any coalition is proportional to its arrival rate, we
 % assume it to be the rate itself.
 The value~$\lambda_C^{\P}$ of each coalition~$C$ must further be
 apportioned between the members of the coalition. Denoting the payoff
 of agent~$i$ by $\phi_i^{\P},$ we therefore have
 $$\sum_{i \in C} \phi_i^{\P} = \lambda_C^{\P} \quad \forall \quad C \in
 \P.$$
 % 
 % Now the providers in each coalition divide the value of coalition
 % among themselves according to some allocation rule, say call it
 % ${\color{red}\{\lambda_{i, C}^{\cal P} \}}$.  Observe here that
 % $ \sum_{i \in C} \lambda_{i, C}^{\cal P} = \lambda_{C}^{\cal P} $
 % for any $C \in {\cal P}$.  {\color{red} values not arrival rates}
 Since the providers are selfish, they are ultimately interested only
 in their individual payoffs. %in their shares $\{\lambda_{i, C}^{\cal
 % P} \}$.
 Thus, the coalition formation between providers is driven by the
 desire of each provider to maximize its payoff, given the statistical
 economies of scale obtained via coalition, and also the
 \emph{constant sum} nature of this game (the sum total of the payoffs
 of all providers equals~$\Lambda$). Thus, the relevant fundamental
 questions are:
\begin{enumerate}
\item Which partitions can emerge as a result of the strategic
  interaction between providers, i.e., which partitions are part of
  stable
  configurations? %Eventually which partition emerges/ is stable?
  Indeed, a precursor to this question is: How does one define a
  natural notion of stability?
\item It is apparent that the answer to the above question hinges on
  how the value of each coalition is divided between its
  members. Thus, the next question is: How is the value of each
  coalition apprortioned between its members in a stable
  configuration?
  % \item It is apparent that the answer to the above question depends
  %   on
  %   the allocation rule/scheduler, which defines the way the value
  %   of
  %   any coaltion is divided among its members. So what kind of
  %   allocation rules are important and which partitions emerge at a
  %   given rule?
  % \item What is the class of allocation rules? An allocation rule
  %   divides the worth of any given coalition among its members in
  %   any given partition.
  %	\item Which coalitions can not be blocked and which is a
  %   stable partition?
  %	% 	
  % \item How is the worth of a coalition divided among its members?
\end{enumerate}
% The answers to these questions are inter-dependent as the blocking
% of coalitions depend on the allocation rule. and the individual
% shares depend on which coalitions are formed.
Our aim in this paper is to answer these questions; such problems can
be studied using tools from cooperative game theory. Note in
particular that the value of any coalition in our formulation depends
on the operational arrangement of agents outside the coalition, i.e.,
on the entire partition.
% This results in a more complicated form of cooperative games called
% partition form games.
This makes the game we study a \emph{partition form game}. In the
remainder of this section, we introduce the notion of a partition
(more precisely, a configuration) being \emph{blocked} by a certain
coalition. These ideas will be used when we define \emph{stable
  configurations} in Section~\ref{stable_config}.
% In the remainder of this section, we define formally the equilibria,
% i.e., stable configurations of this game.

\Cmnt
{\subsection{Existence and Uniqueness of Wardrop Equilibrium: When $n=3$}
  
  At equilibrium, we have
  $pb_1(\lambda_1,N_1) = pb_2(\lambda_2,N_2) =pb_3(\lambda_3,N_3) $.
  Also, we have $\lambda_1 + \lambda_2 + \lambda_3 = \Lambda$.

  \begin{itemize}
  \item When $k_i = N_i \text{ for }  i=\{1,2\}$ then 
    $$
    pb_{12} = \frac{\frac{(a_1+a_2)^{N_1+N_2}}{(N_1+N_2)!}}{\sum_{j=0}^{(N_1+N_2)} \frac{(a_1+a_2)^{j}}{j!}}
    $$ 
    and
    $$
    pb_3 = \frac{\frac{a_3^{N_3}}{N_3!}}{\sum_{j=0}^{N_3} \frac{a_i^{j}}{j!}} \text{ where } a_3 = \frac{\Lambda-(\lambda_1+\lambda_2)}{\mu}
    $$ 
    \subsubsection{Existence and Uniqueness of Wardrop Equilibrium}}
  
  \Cmnt{\textbf{Intuition:}

    We know that the blocking probability of a provider increases with
    increase in the arrival rate to it.  Hence when we calculate
    Wardrop Equilibrium (WE) we have one of the functions as
    increasing while the other one is decreasing. Thus, an
    intersection point will definitely exist. Also, since they are
    strictly increasing or decreasing functions, the solution is
    guaranteed to be unique.
    
    \noindent 
    \textbf{Proof: }
  }
  
%\subsection{Partition-form Coalition-formation Games}

\hide{
  {\bf \st{Allocation rule}} \st{ defines the exact way in which the
    value of any coalition is shared among its members in any given
    partition. Thus, an allocation rule is defined as the following
    vector:}
\begin{eqnarray}
  \Phi &:=&\mbox{\st{$ \left  \{ [ \phi_i^\P ]_i  \right  \}_{\P} , \mbox{ such that, }     \sum_{i \in C} \phi_i^\P = \lambda_C^\P   \text { for all } \, C \in  \P,  $}} \nonumber  \\  && \hspace{40mm} \text{ and for all  }   \P. 
\label{Eqn_allocation_rule}
\end{eqnarray}
\st{Observe here that an allocation rule defines sharing rules for all
  the partitions.  } } %end hide

% We first begin with few definitions.
Given a partition $\P = \{C_1,\cdots,C_k\},$ the set of payoff
vectors consistent with $\P$ is defined as:

\vspace{-4mm}
{\small$${\bm \Phi}^{\P} :=\left  \{\Phi = [\phi_1,\cdots, \phi_n] \in \R^n_+:\ \sum_{j \in C_i} \phi_j = \lambda^{\P}_{C_i}\ \forall\ 1 \leq i \leq k \right \}.$$}
A \emph{configuration} is defined as a tuple~$(\P,\Phi),$ such that
$\Phi \in {\bm \Phi}^{\P}.$ Note that a configuration specifies not just a
partition of the agents into coalitions, but also specifies an
allocation of payoffs within each coalition, that is consistent with
the partition.

% {\bf Payoff-vector} (also known as allocation vector) under any given
% partition $\P$ and any given allocation rule $\Phi$ is defined as
% $\x = \Phi^\P := [ \phi_1^\P, \cdots, \phi_n^\P]$, that satisfies
% $$
% \sum_{i \in C} \phi_i^\P = \lambda_C^\P   \text { for all } \, C \in  \P, 
% $$
% {\color{blue} Observe here that a payoff vector is corresponding to
% a partition.} 

\emph {Blocking   by a coalition:} A configuration
$(\P,\Phi)$ is \emph{blocked} by a coalition~$C \notin \P$ if, for any
partition~$\P'$ containing~$C,$ there exists $\Phi' \in {\bm \Phi}^{\P'}$
such that
%A payoff-vector
%$\x $, corresponding to a configuration $(\P, \Phi^\P)$, is said to be
%blocked by a coalition $C$ (where $C$ is formed by some rearrangement
%of the coalitions of $P$ as described in next section) if for any
%other partition $\P'$ that contains $C$, i.e., if $C \in \P'$, there
%exists a configuration $(\P', \Phi^{\P'})$ such that
\begin{equation}
  \phi_j' > \phi_j \, \forall \, j \in C.
  \label{Eqn_blocking_coalition}
\end{equation}
 Basically, a new coalition can block an existing configuration,
  if each one of its members can derive strictly better payoff from this
  realignment.
Equivalently, $(\P,\Phi)$ is blocked by coalition~$C \notin \P$ if,
for any partition~$\P'$ containing~$C,$
$\lambda^{\P'}_C > \sum_{j \in C} \phi_j.$
Note that the above equivalence hinges on the transferable
  utility assumption inherent in our cooperative game,  by virtue of  which (partial) utilities can be transferred across agents. Intuitively, a
coalition $C \subset \mathcal{N}$ blocks configuration $(\P,\Phi)$, if
the members of $C$ have an incentive to `break' one or more coalitions
of~$\P$ to come together and form a new coalition. In particular, it
is possible to allocate payoffs within the blocking coalition $C$ such
that each member of $C$ achieves a strictly greater payoff,
irrespective of any (potentially retaliatory) rearrangements among
agents outside~$C.$
%
%the coalition can work together independent of other players and
%achieve strictly better utilities for all of them; and they can
%achieve better utilities, no matter in what configuration the rest of
%the players arrange themselves. 
This is referred to in the literature as a \textit{pessimistic
  anticipation rule} (see \cite{pessimistic}, \cite{Shiksha_Perf} and Appendix~A). 

% Finally, we {\bf Stable configuration:} A configuration $(\P, \phi)$
% is stable, if it is not blocked by any coalition $C \notin \P.$

In Section~\ref{stable_config}, we analyse \emph{stable
  configurations}, which are defined as those configurations that
cannot be blocked by a certain broad class of candidate blocking
configurations. Specifically, we consider candidate blocking
coalitions that are formed either via a \emph{merger} of prevailing
coalitions, or via a \emph{split} of a single prevailing
coalition. Also, note that blocking as defined above involves a
revelation of the prevailing payoffs of the agents of the candidate
coalition  $\{\phi_i\}_{i \in C}$. In Section~\ref{stable_config}, we also consider an
alternative definition of blocking, where the `prevailing worth' of
the agents of the candidate blocking coalition is estimated
imprecisely.

Finally, we note that game considered here can also be modelled as a
characteristic form game; the details of this construction are
available in Appendix~A. Indeed, the notion of stable configurations
in the present context is a partition-based generalization of the
classical notion of $\alpha$-core (see \cite{aumann1961},
\cite{alpha-core}), when the characteristic function is defined using
the pessimistic anticipation rule (details are in Appendix~A, see
\eqref{Eqn_pec_bnd}-\eqref{Eqn_pec_nu}).

% This problem can be modelled as the well known characteristic form
% game, the details of which are available in Appendix A.  A partition
% is stable under a given allocation rule if and only if the
% corresponding payoff vector is in $\alpha$-core, when the
% characteristic function is defined using pessimistic anticipation rule
% (details are in Appendix A, see
% \eqref{Eqn_pec_bnd}-\eqref{Eqn_pec_nu}).

% These kind of games can be well modelled by characteristic form games
% as described in Appendix A.  We have the notion of payoff vectors,
% blocking by a coalition and stable payoff vectors as described there.
% We also provided the procedure by which our problem can be modelled as
% a characteristic form game.

% Ours is a partition form game, where we have partitions and our aim
% is to study their stability taking inspiration from the notions
% described in Appendix A. Towards this we associate with each
% partition $\P$, a payoff vector $\Phi^\P$ and talk about the pair of
% configurations $(\P, \Phi^\P)$ that would be stable.

\newcommand{\ulam}{{\underline \lambda}}

\newcommand{\MS}{{Q}}

\section{Stable Configurations}
\label{stable_config}

In this section, we formally define three different notions of stable
configurations, which differ based on the types of candidate blocking
coalitions considered, as well as the precision with which the
`prevailing worth' of the members of the candidate coalition is
estimated. For each of these notions of stability, we characterize the
class of stable configurations. {\it The main takeaway from our
  results is that the interplay between statistical economies of scale
  and the constant sum nature of this game results in configurations
  with three or more coalitions rendered unstable.} In other words,
stable configurations necessarily involve either a duopoly or a
monopoly. Importantly, under all three notions of stability that we
  consider, stable configurations are only composed of such small
  (one/two) sized partitions; however, the payoff vector counterparts
  depend on the particular notion of stability under consideration. 

We begin by defining the different notions of stability we consider.

\subsection{Defining stable configurations}

The first notion of stability we introduce simply restricts the set of
candidate blocking configurations to mergers and splits of prevailing
coalitions. Note that this is a natural restriction from a practical
standpoint, since complex rearrangements between firms in a
marketplace typically arise (over time) from a sequence of mergers and
splits. We refer to this as restricted blocking (RB). Further when one
assumes the precise knowledge of the worth of the blocking candidates,
it leads to the RB-PA (Restricted Blocking--Perfect Anticipation)
rule. We begin with this rule.

{\bf RB-PA rule:} Under this rule, a configuration~$(\P,\Phi)$ is
blocked by a coalition~$Q$ that is formed either via a merger of
coalitions in $\P$ (i.e., $Q = \cup_{C \in \mathcal{M}} C$ for
$\mathcal{M} \subseteq \P$), or via the split of a single coalition in
$\P$ (i.e., $Q \subset C$ for some $C \in \P$), if, for all
partitions~$\P'$ containing $Q,$ there exists
$\Phi' \in {\bm \Phi}^{\P'}$ such that \vspace{-4mm}
$$\hspace{5mm} \phi'_i > \phi_i \quad \forall \quad i \in Q.$$
Equivalently, $Q$ blocks the configuration~$(\P,\Phi)$ if
%$$\lambda^{\P'}_{Q} > \sum_{i \in Q} \phi_i$$ for all partitions~$\P'$
%containing $Q.$ The above condition is equivalent to
\begin{equation}
  \label{eq:blocking_PA}
  \ulam_Q > \sum_{i \in Q} \phi_i, \text{ where }
  \ulam_Q := \min_{\P': Q \in \P'} \lambda^{\P'}_Q.
\end{equation}

A configuration~$(\P,\Phi)$ is stable under the RB-PA rule if it is
not blocked by any merger or split. Note that under the RB-PA rule,
members of a candidate blocking coalition are pessimistic in their
anticipation of the value of the new coalition, in that they consider
`worst case' rearrangements among outside agents. Moreover, it is
possible to allocate the payoff of~$Q$ among its members such that
each member is (strictly) better off, as discussed in previous section.

The next notion we consider uses the same restriction on the set of
candidate blocking configurations, but uses an imprecise estimate of
the prevailing worth of the members of the candidate blocking
configurations in the case of a split, resulting in an imprecise
anticipation of the benefit from the split. We refer to this as the
RB-IA (Restricted Blocking--Imperfect Anticipation) rule.

{\bf RB-IA rule:} Under this rule, a configuration $(\P,\Phi)$ is
blocked by a coalition~$Q$ that is formed by splitting a coalition
$C \in \P$ if: \vspace{-2mm}
\begin{align}
  \label{Eqn_condition_S}
  &\ulam_Q := \min_{\P': Q \in \P'} \lambda^{\P'}_Q > \frac{N_Q}{N_C} \lambda^{\P}_C, \\
  \label{Eqn_condition_S_pt2}
  &\lambda^{\hat{\P}}_Q > \sum_{i \in Q} \phi_i, \text{ where }\hat{\P} = (\P \setminus \{C\}) \cup \{Q, C\setminus Q\}.
\end{align}
Observe here that the right hand side expression  in \eqref{Eqn_condition_S}  is an imprecise estimate of the worth of the  breaking away split, while that in the second equation is the precise value (to be revealed in later part of the negotiations).
On the other hand, under the RB-IA rule, a configuration $(\P,\Phi)$
is blocked by a coalition~$Q$ that is formed by a merger of coalitions
in  $\P$ if \vspace{-3mm}
\begin{equation}
  \label{Eqn_condition_M}
\ulam_Q  >  \sum_{C \subset Q} \lambda^\P_C \mbox{, and } \lambda^{\hat{\P}}_Q > \sum_{i \in Q} \phi_i, 
\end{equation}where $\hat{\P}$ is   the new partition after the merger.  
Observe here    that  $\sum_{i \in Q} \phi_i=  \sum_{C \subset Q} \lambda^\P_C$ and hence the second condition is immediately satisfied for merger, because $\ulam_Q  \le \lambda^{\hat{\P}}_Q$. 
Finally, {\it a configuration is
stable under the RB-IA rule if it is not blocked by any merger or
split.}

Note that RB-PA and RB-IA differ only in the condition for blocking
due to a split. This is natural, since the net worth of coalitions
 $\{\lambda_C^\P\}_{C \in \P}$ is often common knowledge,
whereas the internal payoff allocation within a coalition can often be
confidential. Let us therefore interpret the condition for blocking
due to a split under RB-IA. Condition~\eqref{Eqn_condition_S} can be
interpreted as a first stage check on the feasibility of the split, by
(imperfectly) estimating the total prevailing worth of the members
of~$Q$ as proportional to their contribution to the service capacity
within $C.$ On the other hand, the
condition~\eqref{Eqn_condition_S_pt2} can be interpreted as the final
stage check on split feasibility, that ensures that it is possible to
allocate the payoff of~$Q$ among its members such that each member is
(strictly) better off from the split.

Finally, we consider the stability notion resulting from the most
general model for blocking. Here, we allow blocking by an arbitrary
coalition (which  also includes merger of partial splits), with a precise estimation of the prevailing worth of the
members of the blocking coalition. We refer to this as the GB-PA
(General Blocking--Perfect Anticipation) rule.

{\bf GB-PA rule:} Under this rule, a configuration~$(\P,\Phi)$ is
blocked by \emph{any} coalition~$Q \notin \P$ if
\eqref{eq:blocking_PA} holds. A configuration is stable under the
GB-PA rule if it is not blocked by any coalition.

Clearly, the set of stable configurations under the GB-PA rule is a
subset of the set of stable configurations under the RB-PA rule.

Having defined our notions of stability, we now consider each notion
separately, and characterize the resulting stable configurations. We
begin with RB-IA, which (it turns out), admits the broadest class of
stable configurations.

\subsection{Stable configurations under RB-IA}

Our first main result is that all configurations involving partitions
of size three or more are unstable. In other words, only monopolies or
duopolies can be stable.

\begin{theorem}
\label{Thm_duo_mono}
\textit{Under the RB-IA rule, any configuration~$(\P,\Phi)$ with $|\P| \geq 3$ is not stable.}
\end{theorem}
\noindent
\textbf{Proof: } See Appendix C. \eop

Intuitively, partitions of size three or more are unstable because of
the statistical economies of scale resulting from a merger (see
Statement~$(ii)$ of Theorem \ref{Thm_WE}). Specifically, if
$|\P| = k \geq 3,$ it can be shown that any merger between $k-1$
coalitions in~$\P$ would block the configuration~$(\P,\Phi).$

Having ruled out the possibility of stable configurations with three
or more coalitions, we now explore the two remaining possibilities:
stable configurations involving the grand coalition, and those involving 2-partitions.

\noindent {\bf Grand Coalition:} Defining $\P_G := {\cal N}$ as the
grand coalition, it is clear that any configuration of the
form~$(\P_G,\Phi)$ can only be blocked by a split. We now show that
unless a single agent owns more than half the total service capacity
of the system, such a block is always possible. In other words, any
configuration involving the grand coalition is unstable unless there
is a single `dominant' agent. On the other hand, if there is a single
agent who owns more than half the service capacity, we show that {\it there
exists stable configurations of the form $(\P_G,\Phi),$ i.e., the
grand coalition can be stable in the presence of single dominant agent.}

% The grand coalition (GC) $\P^G := {\cal N}$ can't have a successful
% merger as in \eqref{Eqn_condition_M} for obvious reasons.  Any
% 2-partition can't merge successfully into GC, because the total
% utility after merger equals that before merger, $\Lambda$. This
% constant sum nature of the game instrumental in the following
% impossibility result.  In other words, GC can not result from a
% successful merger of any of the existing partitions, one can
% encounter GC only when the system starts with a GC.  Now consider
% that the system starts with GC. In the immediate following we show
% that the GC is not stable, irrespective of the starting allocation
% vector.

\begin{theorem}
  \label{Thm_GC} 
 \textit{Under the RB-IA rule:}
  
  \noindent \textit{i) If $N_1 \leq \sum_{i \in \mathcal{N}; i \neq 1}N_i$,
  there exists no payoff vector $\Phi$ consistent with ${\P}_G$, such
  that $({\P}_G, \Phi)$ is stable.}

  \noindent \textit{ii) If $N_1 > \sum_{i \in \mathcal{N}; i \neq 1}N_i$, there
  exists atleast one payoff vector $\Phi$ consistent with ${\P}_G$, such
  that $({\P}_G, \Phi)$ is stable.}
\end{theorem}
\textbf{Proof:} See Appendix C. \eop

\Cmnt{ ii) {\color{green} GC stable under some allocation vectors:
    when the players are asymmetric and there exist no subset
    $S \subset \mathcal{N}$ such that $N_1 < \sum_{i \in S}N_i$}}

\noindent {\bf Two-partitions:} We finally turn to configurations
involving 2-partitions. Two-partitions can, without loss of
generality, be represented as $\P = \{C_1,C_2\},$ with
$N_{C_1} \geq N_{C_2}.$ We now show that under the RB-IA rule, the
stability/instability of a configuration~$(\P,\Phi)$ depends majorly
on the value of $k:=N_{C_1}.$ Interestingly, the stability/instability
of a configuration is  generally not  influenced by the associated
payoff vector~$\Phi$ under the RB-IA rule. (This is not true under
RB-PA or GB-PA rules.)

% {\color{red}Let $N := \sum_i N_i$ be the total number of servers of
% all the players and} assume $N_1 \ge N_2 \ge \cdots N_n$, without
% loss of generality.

% Any 2-partition $\P = \{C_1, C_2\}$ can be identified uniquely by
% $k := N_{C_1} $ the number of servers of its first coalition.  For
% simpler notations, we let $\lambda_k := \lambda_{C}^\P$, when
% $N_C = k$, and this definition is possible because all the servers
% are identical.  Our aim now is to find the values of $k$ that
% renders the 2-partition stable.

Formally, let~$\lambda_k := \lambda_{C_1}^\P.$ Note that by
Theorem~\ref{Thm_WE}, $\lambda_k$ is the unique zero of the following
function of $\lambda$ (see \eqref{Eqn_WE_properties}):
$$h(\lambda) := \frac { \lambda^k } { k! }  \sum_{j=0}^{N-k} \frac{
  (\Lambda-\lambda)^j }{j!}  -
\frac{{(\Lambda-\lambda)}^{N-k}}{(N-k)!}  \sum_{j=0}^{k}
\frac{\lambda^j} {j!}.$$
% Observe here that $\lambda_{N-k} = \Lambda - \lambda_k$. 
Next, define $\Psi(k; \Lambda):= \lambda_k/k$ as the utilization per
server of the larger coalition. Finally, define
\begin{eqnarray}
  k^* (\Lambda) := \argmax_{k: k = N_{C_1}} \Psi(k ; \Lambda). \label{Eqn_kstar}
\end{eqnarray}
Note that $k^*(\Lambda)$ is the set of values of $k$ that maximizes
the per-server utilization of the larger coalition. 

Let
${\mathbb C}^* := \{ C \subset {\cal N}: N_C \in k^* (\Lambda) \} $ be
the set of coalitions $C$,   that can derive maximum per-server utilization,  irrespective of the operational arrangement of  the other agents.  In the
following lemma, we provide a sufficient condition for a class of
configurations to be stable. We adopt the following {\it convention: A
partition $\P$ is stable if all configurations involving it are
stable, i.e., configuration $(\P, \Phi)$ is stable for any
$\Phi \in {\bm \Phi}^\P.$} 

\begin{lemma}
\label{Lemma_stable_two_RB-IA}
  % {\bf [Stable configuration under rule-A]} 
  \textit{Consider the RB-IA rule. A 2-partition $\P = \{C_1,C_2\}$ is stable
  if there exists no coalition $S \subset C_i \text{ for }
  i=\{1,2\}$ such that: \vspace{-6mm}
 $$ \hspace{24mm} \frac{ \ulam_S } {N_S } > \frac{\ulam_{C_i}}{N_{C_i}}= \frac{\lambda_{C_i}^{\P}}{N_{C_i}}. \hspace{24mm} \mbox{ \eop}$$}
\end{lemma}

The proof of the lemma follows directly from the definition of
stability. A consequence of this lemma is the following.

\begin{theorem}
  % {\bf [Stable configuration under rule-A]} 
  \label{Thm_two_partition}
  \textit{Consider the RB-IA rule. Any 2-partition $\P = \{C_1,C_2\}$ with one
  of the coalitions from $\mathbb {C}^*$ is a stable
  partition. Additionally, any partition $\P = \{C_1,C_2\}$ satisfying
  $N_{C_1} = N_{C_2} = N/2$ is stable.}
\label{Thm_Prule_stability}
\end{theorem} {\bf Proof:} See Appendix C. \eop

Note that under the RB-IA rule, we have identified a class of
\emph{stable partitions}, i.e., these partitions are stable for any
consistent payoff vector. In Section~\ref{sec_two_partition}, we provide a complete
characterization of the class of stable partitions under RB-IA, in the
heavy and light traffic regimes.

\subsection{Stable configurations under RB-PA}

\hide{
The equations \eqref{Eqn_condition_M}-\eqref{Eqn_condition_S} can be
seen as the initial anticipation/assessment by the new coalition
before plunging into the action.  They consider pessimal rule for
anticipation to safeguard themselves from future mergers/coalitions.
The merger/split will really be through if further equation
\eqref{Eqn_Example_fair_rule_M} and \eqref{Eqn_Example_fair_rule_S} is
satisfied, and one can view this as final assessment after further
fact-checking.

Alternatively one can consider one step anticipation rule, in which a
new coalition $Q$ (merger/split) is formed if the following is
satisfied:
\begin{eqnarray}
\label{Eqn_alternate_rule}
\ulam_Q  >  \sum_{i \in Q}  \phi_i^\P  .
\end{eqnarray}
Note by definition of $\ulam_Q$, under this anticipation rule, the new
coalition always perform superior (as $\ulam_Q \le \lambda_Q^{\P'} $
for all $\P'$) and hence can be seen as one step assessment.
} %end hide

Next, we consider stable configurations under the RB-PA rule, which
presents interesting contrasts to the RB-IA rule. Under this rule, 
we show that only coalitions involving 2-partitions can be stable,
i.e., configurations involving the grand coalition, or
involving~$k$-partitions with $k \geq 3$ are always
unstable. Moreover, the stability/instability of configurations
involving 2-partitions depends on the associated payoff vector.

To demonstrate this, we define a special \emph{proportional} payoff
vector, where the value of each coalition is divided between its
members in proportion to the number of servers they bring to the
coalition. Formally, the proportional payoff vector $\Phi^{\P}_p$
associated with a partition~$\P$ is defined by:
% We say the payoff vector $\Phi^{\P}_p$ to be \textit{proportional} if
% the share of value of any coalition (in any partition) to any of its
% members is proportional to value of its resources divided by the total
% value of the resources of all the members of the coalition. For
% example, in our queueing model:
\begin{eqnarray}
\label{Eqn_PSA}
\phi^{\P}_{p, i}  =  \frac{N_i}{ \sum_{j \in C} N_j }   \lambda_C^\P \mbox{ for any } i  \in C \in \P.
\end{eqnarray}

Our results for the RB-PA rule are summarized as follows.
\begin{theorem}
  % {\bf [(Un)Stable configurations under rule-B]}
  \label{Thm_stable_config_ruleB}
 \textit{ Under the RB-PA rule:}

  \noindent \textit{i) No configuration involving the grand coalition is
  stable.}

  \noindent \textit{ii) No configurations involving $k$-partitions, where
  $k \geq 3$ are stable.}

  \noindent \textit{iii) For any 2-partition $\P = \{C_1,C_2\},$ where one of
  the coalitions lies in $\mathbb{C}^*,$ $(\P,\Phi_p^\P)$ is
  stable.}
  
    \noindent \textit{iv) For any 2-partition $\P = \{C_1,C_2\},$ where  $N_{C_1} =N_{C_2} = N/2$,  $(\P,\Phi_p^\P)$ is
  stable.}
  
 \noindent \textit{v) More generally, consider any
    2-partition $\P = \{C_1,C_2\}$, that is stable under RB-IA
    rule. Then $(\P,\Phi_p^\P)$ is stable under RB-PA rule. Further
    there exists a neighbourhood ${\cal B}^\P_p$ of the payoff vector
    $\Phi_p^\P$ such that $(\P,\Phi)$ is stable for all
    $\Phi \in {\cal B}^\P_p$.}
  
  % under rule-B, where $\P = \{S, \mathcal{N}\backslash S\}$ is
  % any stable 2-partition of Theorem \ref{Thm_two_partition} and
  % $\Phi_p^\P$ is the PF payoff vector corresponding to $\P$ defined
  % by
  % \eqref{Eqn_PSA}.
 \label{Thm_R_rule_stability}
\end{theorem}
\textbf{Proof:} See Appendix C. \eop

Theorem~\ref{Thm_stable_config_ruleB} conveys that partitions that are
stable under the RB-IA rule (irrespective of the associated payoff
vector), are also part of stable configurations under RB-PA, but under
a restricted class of payoff vectors. Specifically, the payoff vectors
we identify are `close' to proportional allocations. Whether there are
other natural payoff structures  that also induce stability under RB-PA, is an
interesting question for future work. 

%%%%%%%%%%%%%%%%%%%%%%%%%%%%%%%%%%%%%%%%%%%%%%%%%%%
%%%%%%%%%%%%%%%%%%%%%%%%%%%%%%%%%%%%%%%%%%%%%%%%%%%
%%%%%%%%%%%%%%%%%%%%%%%%%%%%%%%%%%%%%%%%%%%%%%%%%%%

\hide{
  
{\color{blue} In any real scenario, agents consider joining the
  existing collaborations or splitting from the existing ones.
  Further the payoffs derived by the agents after the new
  collaborations, depend upon the previous payoffs and the influence
  of the new operational arrangement. We consider this aspect; we
  basically consider history based (in particular Markovian)
  allocation rules.  In this section our allocation rules are given by
  $\Phi^{\P'} = f( \Phi, \P, \P')$ for some function $f$, where $\Phi$
  is the old allocation vector and $\P$, $\P'$ are respectively the
  old and the new partitions.  The new allocation depends upon the
  utilities $\{ \lambda_C^\P\}_{C \in \P}$ and
  $\{ \lambda_S^{\P'}\}_{S \in \P'}$
  ${\color{green} \{\lambda_{\MS}\}_{\MS \in \P'}}$ derived by the
  operational units in both the partitions.  }
} %end hide

%end hide
%%%%%%%%%%%%%%%%%%%%%%%%%%%%%%%%%%%%%%%%%%%%%%%%%%%
%%%%%%%%%%%%%%%%%%%%%%%%%%%%%%%%%%%%%%%%%%%%%%%%%%%
%%%%%%%%%%%%%%%%%%%%%%%%%%%%%%%%%%%%%%%%%%%%%%%%%%%

\Cmnt{\subsection{Two partitions}  We are now left with 2-partitions or partitions of size  2.  
In this section we derive the partitions among  the 2-partitions  that are  possibly stable. We also discuss the allocation vectors that render them stable.  

Let $N := \sum_i N_i$ be the total number of servers of all the  players and assume  $N_1 \ge N_2 \ge \cdots N_n$, without loss of generality.  Any 2-partition $\P = \{C_1, C_2\}$ can be  identified uniquely by $k := |C_1| $ the number of servers  of its  first coalition.   For simpler notations,  we  let $\lambda_k := \lambda_{C}^\P$, when $|C| = k$,  and this definition is possible because all the servers are identical. 
Our aim now  is to find  the values of  $k$ that renders the 2-partition stable. 

Let $\Psi(k ; \Lambda)$ represent $\lambda_k / k$ for any given $\Lambda$. By Theorem \ref{Thm_WE},   the function $\Psi(k ; \Lambda)$  equals $1/k$ times the unique  zero  of the following function  of $\lambda$ (see \eqref{Eqn_WE_properties}):
$$
h(\lambda) :=   \frac { \lambda^k } { k! }   \sum_{j=0}^{N-k} \frac{ (\Lambda-\lambda)^j }{j!}   -  \frac{{(\Lambda-\lambda)}^{N-k}}{(N-k)!}     \sum_{j=0}^{k} \frac{\lambda^j} {j!}.
 $$
 Observe here that $\lambda_{N-k} = \Lambda - \lambda_k$. 

 We immediately have the following:
 \begin{theorem}
{\bf [Stable partition]} 
\label{Thm_two_partition}
i) The value of $k$  that renders the  partition $\P = \{S,  {\cal N}\backslash S\}$ (with $|S| = k$)   stable  is given  by:
\begin{eqnarray}
 k^* (\Lambda) := \arg \max_{k: k = |C|, C \subset {\cal N} }   \Psi(k ; \Lambda). \label{Eqn_kstar}
\end{eqnarray}

ii) If in any 2-partition $\P = \{C_1,C_2\}$, there exists no coalition $S \subset C_i \text{ for } i=\{1,2\}$ such that equation \eqref{Eqn_condition_S} is satisfied, then such a 2-partition is also stable.

Both the results are true irrespective of the initial payoff vector $\Phi^\P.$
\end{theorem}
{\bf Proof:} See Appendix B. \eop

\subsection{Alternate  merger/split rules}  The equations \eqref{Eqn_condition_M}-\eqref{Eqn_condition_S} can be seen as the  initial anticipation/assessment  by the new coalition before plunging into the action.  They consider pessimal rule for anticipation to safeguard themselves from future mergers/coalitions.  The merger/split will really  be through if further equation \eqref{Eqn_Example_fair_rule_M} and \eqref{Eqn_Example_fair_rule_S} is satisfied, and one can view this as final assessment after further fact-checking.  

Alternatively one can consider one step anticipation rule, in which a new  coalition $Q$ (merger/split) is formed if the following is satisfied:
\begin{eqnarray}
\label{Eqn_alternate_rule}
\ulam_Q  >  \sum_{i \in Q}  \Phi_i^\P  .
\end{eqnarray}
Note by definition of $\ulam_Q$, under this anticipation rule, the new coalition always perform superior (as $\ulam_Q \le \lambda_Q^{\P'} $ for all $\P'$) and hence can be seen as one step assessment.  Interestingly under this condition, there exists no stable coalition:
\begin{theorem}
i)  There exist no (initial) allocation vector under anticipation rule \eqref{Eqn_alternate_rule} that  renders GC stable.

ii) The value of $k$ that renders the partition $\P = \{S, \mathcal{N},S\}$ with $|S| = k$ stable is given by:
\begin{eqnarray}
 k^* (\Lambda) := \arg \max_{k: k = |C|, C \subset {\cal N} }   \Psi(k ; \Lambda). \label{Eqn_kstar_R}
\end{eqnarray}
when the initial allocation vector is same as proportional share.

iii) If in any 2-partition $\P = \{C_1,C_2\}$, there exists no coalition $S \subset C_i \text{ for } i=\{1,2\}$ such that equation \eqref{Eqn_condition_S} is satisfied and with initial allocation vector same as proportional share, then such a 2-partition is also stable.
%All the partitions are unstable irrespective of the allocation vectors under the anticipation rule \eqref{Eqn_alternate_rule}.
\label{Thm_R_rule_stability}
\end{theorem}

\textbf{Proof:} See Appendix B. \eop}

\subsection{Stable configurations under GB-PA}

Finally, we consider the GB-PA rule, which allows for a configuration
to be blocked by any arbitrary coalition, formed via (possibly)
simultaneous partial splits and partial mergers. This is also the case
in the general definition of blocking by a coalition as given in
Section~\ref{sec_model}, as well as in \cite{Shiksha_Perf} and Appendix A. We briefly discuss
this case here; a complete characterization of stability under GB-PA
will be pursued as future work.
 
It is immediate that since pure mergers or pure splits are included
within the class of candidate blocking coalitions under GB-PA,
\emph{any configuration that is not stable under RB-PA is also not
  stable under GB-PA.} From Theorem \ref{Thm_stable_config_ruleB}, it
therefore follows that the grand coalition, and $k$-partitions with
$k \geq 3,$ can never be part of a stable configuration under GB-PA.

% in this general move and hence a configuration not stable in the
% previous sub-sections will also be not stable for this general move;
% thus all the unstable configurations of Theorem
% \ref{Thm_stable_config_ruleB} are also also not stable with general
% move under rule-B.

Now, consider the following stable configuration (under the RB-PA
rule) from the same theorem, $(\P, \Phi_p^\P)$ with
$\P = \{C, {\cal N} \setminus C\},$ and $C \in \mathbb{C}^*.$ If there
exists $S \subset C$ such that
$C' :=   S \cup ({\cal N} \setminus C)\} \in \mathbb{C}^*,$ then
$C'$ would block the configuration $(\P, \Phi_p^\P),$ since 

\vspace{-4mm}
{\small \begin{align*}
  \sum_{i \in C'} \phi_{p,i}^\P & = \sum_{i \in S} \phi_{p, i}^\P + \sum_{i \in {\cal N} \setminus C} \phi_{p, i}^\P
                                  = N_S \frac{\lambda^\P_{C}}{N_C} + N_{{\cal N} \setminus C} \frac{\lambda^\P_{{\cal N} \setminus C}}{N_{{\cal N} \setminus C}}   \\
                                  & < N_S \frac{\lambda^\P_{C}}{N_C} + N_{{\cal N} \setminus C} \frac{\lambda^\P_{C}}{N_{ C}}  
   =  (N_{C'})  \frac{\lambda^{\P'}_{C'}}{N_{C'}}  =  \ulam_{C'},
\end{align*}}%
where~$\P' = \{C', \mathcal{N} \setminus C'\}.$ This suggests that
stability under GB-PA is more \emph{fragile} as compared to (the
arguably more practical) RB-PA rule. It would thus be interesting to
investigate the following questions in the future: a) Do there exist
stable configurations under GB-PA (i.e., under the broadest class of
blocking coalitions)?  b) In a dynamic game environment (of the kind
considered in Section~\ref{sec:dynamic}), would GB-PA induce a limit
cycle between certain equivalent configurations?
% would the dynamics toggle between few equivalent configurations,
% etc.

\hide{This can be blocked by coalition where $S$ is an appropriate
  subset of $C^*$ such that $N_C = k^*$ (if possible) because then,
\begin{eqnarray*}
   \sum_{i \in C} \phi_{p, i}^\P & = & \sum_{i \in S} \phi_{p, i}^\P + \sum_{i \in {\cal N} \backslash C^* } \phi_{p, i}^\P = N_S \frac{\lambda_{C^*}}{k^*} + \frac{\lambda_{{\cal N}-C^*}}{n-k^*}   \\
   &< &
   (N_S + n-k^*)  \frac{\lambda^\P_{C^*}}{k^*}  =  \ulam_{C}.
\end{eqnarray*}
}

\section{Dynamic coalition formation game}
\label{sec:dynamic}
\Cmnt{
{\color {red} will be removed later

  We have two assessment/anticipation rules when a move (merger/split)
  from an unstable configuration is considered as in the previous
  section, and we summarize the same in the following:

$\bullet$ {\bf Anticipation rule (RB-IA rule} - The merger satisfies
\eqref{Eqn_condition_M} and one can always devise a new payoff vector
which is strictly beneficial for all in the new coalition, i.e.,
satisfies \eqref{Eqn_fair}. Alternatively a split is possible if
\eqref{Eqn_condition_S} is satisfied, however the split move is
realized only if there exists a payoff vector that satisfies
\eqref{Eqn_fair}.

Restricted blocking with imperfect information anticipation (RB-IA)
rule.

$\bullet$ {\bf Anticipation rule, RB-PA rule } - The merger/split
directly satisfies \eqref{Eqn_alternate_rule}, and then move can
always be realized successfully.  Restricted blocking with Perfect
information anticipation (RB-PA) rule. }}

In this section we consider a dynamic version of the game discussed in
the previous sections. We begin with a queueing system and agents
operating in some configuration.  The agents are constantly on the
lookout for greener pastures, and would stop their
quest only if they are satisfied with the existing configuration.

Agents may consider joining existing collaborations or may consider
splitting from some of them.  The (new) payoffs derived by the agents
after the new collaborations (if any), depend upon the previous
payoffs and the value of the new operational
arrangement/coalition. Depending upon the new payoffs, some of
the agents might again consider another movement.  On the other hand,
the system might settle, if all the agents are satisfied with the
configuration.  We study this aspect by considering a sequence of
dynamic coalition formations.

%We consider this aspect by  considering  history based (in particular Markovian) allocation rules. 
%Our allocation rules of this section are given by $\Phi^{\P'} = f( \Phi, \P, \P')$ for some function $f$, where $\Phi$ is the old  payoff  vector and  $\P$, $\P'$ are respectively the old and  the new partitions. 
%The new payoff vector depends upon the  utilities $\{ \lambda_C^\P\}_{C \in \P}$  and $\{ \lambda_S^{\P'}\}_{S \in \P'}$ ${\color{green} \{\lambda_{\MS}\}_{\MS \in \P'}}$ derived by the operational units in both the partitions. 

{\bf Dynamics:} The system starts with some operational
arrangement given by $\P_0$ and with a payoff vector
$\Phi^{0} = [ \phi_1^{0}, \cdots, \phi_n^0]$. If the configuration
$(\P_0, \Phi^{0})$ is stable as defined in previous sections, it is
not beneficial for any member to consider any (coalitional) deviation
and hence the system does not undergo any change.  If that is not the
case, some members of the partition merge/split.
 
There could be more than one movement (merger/split)  that may be successful,  under both the assessment
rules (RB-PA and RB-IA). We assume that
  any such blocking coalition~$Q$ is equally likely to form, causing
  the system to evolve to a new partition, say $\P_1.$ In case of the
RB-IA rule, any new payoff vector $\Phi^{1}$ that satisfies
 $\phi^1_i > \phi^0_i$ for all $i \in Q$ would
suffice.  We discuss the RB-PA rule towards the end of this section.

The system stops if the new configuration $(\P_1, \Phi^{1})$ is
stable.  If not, it switches to yet another configuration
$(\P_2, \Phi^2)$ randomly (and equally likely among all possible
movements) in a similar way. This  evolution continues until
stopped by a stable configuration. Our aim is to understand if such a
limit stable configuration exists.

By the results of the previous section we have stable configurations
only with 2-partitions or grand coalition and we immediately have the
following result under the following assumption:

 \begin{figure}[h]
 \vspace{-4mm}
 \centering
% \begin{minipage}{.5\textwidth}
   \centering 
    \includegraphics[width=.8\linewidth,height=4cm]{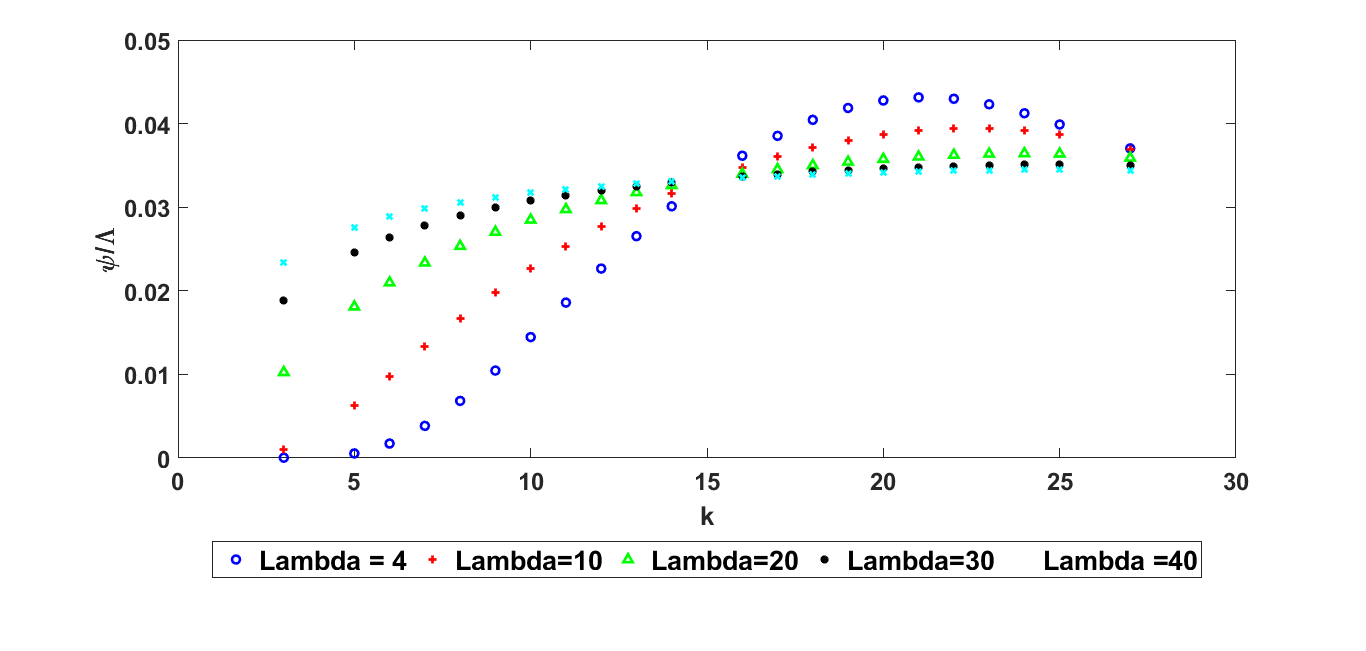}
    \vspace{-4mm}
    \caption{$\Psi(k;\Lambda)/\Lambda$ v/s $k,$  with $[N_i] = [9, 7, 6, 5, 3].$ 
      \label{fig:psi_vs_k}}
   \vspace{-3mm}
 %\end{minipage}%
 \end{figure}

{\bf A.1)} %  A 2-partition $\P  = \{C_1, C_2 \}$  is stable under RB-IA rule, if and only if  one of its coalitions is  from $\mathbb{C}^*$ or 
%
 %
% Let $C^*$ represent any coalition with  $N_{C^*} = k^*$, given in Theorem \ref{Thm_two_partition} and let ${\cal C}^*$ be the collection of all such $C^*$. 
%If there exists any coalition with a strict subset that improves in the PF sense,  then it must include one of the coalitions $C^*$  
If $C$ is any coalition that does not contain any element of  ${\mathbb C}^*$,
i.e., if $C \cap C^* \ne C^*$ for all $C^* \in {\mathbb C}^*$, then we
have the following:
$$
\frac{ \ulam_S }{N_S}  < \frac{ \ulam_C }{N_C}  \mbox{ for all strict subtsets,  }  S \subset C.
$$

Basically this assumption ensures that any 2-partition that is not stable, necessarily contains a coalition that is a strict superset of an element from ${\mathbb C}^*$.  From simulations, we
have seen that this assumption is satisfied by our queuing system for
all the cases that we considered (for example, see
Figure~\ref{fig:psi_vs_k}) and further by~Theorem \ref{Thm_approx} and
Lemma~\ref{low_traffic}  can be shown to hold  under heavy
and light traffic conditions. Under this assumption, we can show that
the dynamics stops after finite number of movements.
\begin{theorem} {\bf [Convergence]}
  Assume {\bf A}.1. Then the random dynamics under RB-IA rule
  converges to one of the stable configurations under RB-IA rule in
  finite number of steps, with probability one.
\end{theorem}
{\bf Proof:} 
We first show that
 starting from any $k$-partition $\P$ with $k > 2$, the dynamics hits a 2-partition with probability one: 
  i) from any such $\P$, there exists at least one direct path to  a 2-partition with probability strictly greater than zero, as given in the proof of Theorem \ref{Thm_duo_mono}; ii) thus  there exists a non-zero uniform lower bound  ${\underline p} >0$ on  the probability of hitting a 2-partition,  irrespective of  the starting $k$-partition,  because of finitely many such  partitions; and  iii) thus by independence,  the dynamics hits a 2-partition with probability one in finite number of steps (uniformly upper bounded by a geometric random variable with parameter  ${\underline p}$). 
  
  Similarly starting from   the grand coalition the system either evolves to a 2-partition or stops. 

If the dynamics hits one of the stable  partitions (among $2$-partitions), we are done. If not, by {\bf A.}1,  the 2-partition (say $\P = \{C, {\cal N}   \backslash  C\}$)  is such that (without loss of generality)   $N_C> k^*$ and $C$ contains a $C^* \in {\cal C}^*$.   The movement from $\P$ to $\P_1  := \{C^*,  C  \backslash  C^*,    {\cal N}  \backslash  C\}$ is possible by \eqref{Eqn_condition_S}  because clearly  by definition of $k^*$ and $C^*$
$$
\frac{ \ulam_{C^*} }{k^*}  >   \frac{ \ulam_C}{N_C}  =  \frac{ \lambda_C^\P } {N_C} \mbox{ which implies }    \ulam_{C^*} >   \lambda_C^\P  \frac{k^*} {N_C}.
$$
From $\P_1$ merger of $ C  \backslash  C^*$ and   $  {\cal N}  \backslash  C$ to $\P_2  := \{C^*, {\cal N} \backslash C^* \}$  is possible  by \eqref{Eqn_condition_M}, as clearly 
$$
\ulam_{ {\cal N}  \backslash  C^*}  >  \lambda^{\P_1}_{C  \backslash  C^*} +   \lambda^{\P_1}_{{\cal N}  \backslash C^*},
$$as in the proof of Theorem \ref{Thm_duo_mono}. The succession of these two events occur with  probability that can be lower bounded by a strictly positive number ${\underline p}'$, uniformly across all such starting 2-partitions.  As in the previous paragraph, any upward movement will return to a 2-partition with probability one and in each of these returns there is  uniform lower bound ${\underline p}'$ on the probability of return to the stable 2-partition with a $C^*$. Hence  the theorem.  \eop

\Cmnt{ {\color{red} I think we can get rid of A.1, because of the
    following: If there is any 2-partition $\P = \{C_1, C_2\}$ which
    contain two or one coalitions (say $C_i$) that contain a strict
    subset (say $S_i$) that strictly improve in PF sense. Then one
    guarantee the existence of a further subset $S'_i \subset S_i$ in
    each of them which does not have further subsets that strictly
    improve. Then one can consider two upward movements followed by
    suitable downward movements (using the fact that 3-partition is
    not stable). If the last part is done, it would be done.  }}

The above theorem proves that the random dynamics under RB-IA rule is
stopped in finite number of steps with probability one, and the limit
is a stable partition.
However under this imprecise anticipation rule, it is important to observe that the payoff vector at the stopped configuration can be  arbitrarily skewed (as also indicated in Theorem \ref{Thm_two_partition}).  

\subsection*{Dynamics under RB-PA rule}
Under RB-PA rule, the random dynamics behaves exactly similar to RB-IA
rule (as described in the proof), however it may not stop even after
touching a stable 2-partition, stable under RB-IA. As seen from Theorem
\ref{Thm_stable_config_ruleB} for RB-PA rule, the payoff vector is
equally important in the definition of stable configuration.

This shows the importance of appropriate reallocation of individual
shares after the new move towards the stability of the new system; it
is not sufficient to only ensure all members of the new
coalition derive positive increments, rather we will require that the
new allocation matches the payoff vector in the corresponding stable
configuration.  As seen from Theorem \ref{Thm_stable_config_ruleB},
one of the payoff vectors that provides stable configurations is the proportional  payoff vector given by equation \eqref{Eqn_PSA}.  Thus one probably has to design reallocation
policies that converge towards the proportional payoff vectors for the dynamics
under the RB-PA rule to stop.

Alternatively there might be other payoff vectors which would also
form a part of the stable configurations and they could be the ones at
limit.  We would study this aspect in the future, but for now we could
say that one can't have partitions of size greater than 2 or the grand
coalition (when none of the agents dominate) to be a part of the limit
(stable) configuration (if one exists), in view of Theorems
\ref{Thm_duo_mono} and \ref{Thm_GC}.  We can also say that the
dynamics stops if it hits upon a configuration with stable 2-partition
and the corresponding proportional payoff vector \eqref{Eqn_PSA}.

\Proofs{
\subsection*{Dynamics under GB-PA rule} We only have preliminary
results for this rule. As mentioned before, all the configurations
that we discussed before are not stable.  It is not difficult to show
that the dynamics does not stop  even if it starts
with  or hits a stable configuration under RB-PA rule identified in Theorem \ref{Thm_R_rule_stability}.  It is interesting
to observe that the dynamics toggles between stable configurations of
RB-PA rule, even when it starts with one of them.
}{}

 %{\color{green} History dependent and independent partitions}

\vspace{-4mm}
\section{Stable 2-partitions}
\label{sec_two_partition}

We now obtain the stable  partitions  of Theorems \ref{Thm_two_partition} (RB-IA) and \ref{Thm_stable_config_ruleB} (RB-PA). We achieve this by considering heavy and light traffic regimes. 
 Recall any  2-partition $\P = \{C_1, C_2\}$ can be  identified uniquely by $k := N_{C_1} $,  $\lambda_k := \ulam_{C_1}$, when one considers optimizing   $\Psi(k ; \Lambda) = \lambda_k / k$.
We first begin with analysis of $k^*$ defined in \eqref{Eqn_kstar}. 
%We also derive the other stable partitions of Theorems \ref{Thm_two_partition} and \ref{Thm_stable_config_ruleB} in heavy traffic regime. 

	\subsection{Heavy Traffic} 

Our aim in this section is to derive the analysis using some appropriate approximations, and then prove that the derived results are valid for for all arrival rates  with	$\Lambda > {\bar \Lambda}$, where
$ {\bar \Lambda}$  is a big enough value.  %We let $\lambda_1, \lambda_2 $ respectively  represent $\ulam_{C_1}$ and $\ulam_{C_2} $.
 
\Proofs{

	The WE can also be obtained by equating the reciprocal of blocking probabilities \eqref{Eqn_PB}  and the first order approximation suggests that approximate WE can be obtained by  solving the following equation 
	written in terms of $\lambda_1$ and $\lambda_2 := \Lambda-\lambda_1$,
	\begin{eqnarray*}
	1+\frac{k}{\lambda_1}  =   1+\frac{N-k}{\lambda_2}  \mbox{ and then the solution 
	}
 \lambda_1^*  =  \frac{k}{N} \Lambda.
	\end{eqnarray*}
	Thus with this approximation the share of any agent    $i \in C_1$ under  proportional payoffs \eqref{Eqn_PSA}  equals $ N_i / k    *  k / N  \Lambda = N_i / N \Lambda  $, irrespective of $k$.  Thus this approximation is not sufficient and we now consider the second order approximation under which we require zeros of the following: 
	\begin{eqnarray}
	1+\frac{k}{\lambda_1}+ \frac{k(k-1)}{\lambda_1^2} =  1+\frac{N-k}{\lambda_2}+\frac{(N-k)(N-k-1)}{\lambda_2^2} ,\nonumber 
	\end{eqnarray}
	which after some simple calculations leads  to the following fixed point equation (of $\lambda_1 \in [0, \Lambda]$ and $\lambda_2 := \Lambda-\lambda_1$):
	\begin{equation}
	 {\frac{\lambda_1}{k}= \frac{\lambda_2}{N-k} \Bigg(\frac{ 1+\frac{k}{\lambda_1}-\frac{1}{\lambda_1} }{  1+\frac{N-k}{\lambda_2}-\frac{1}{\lambda_2} } \Bigg)} .
	\label{Eqn_fixed_point_eq}
	\end{equation}
	Let $\psi (k) :=  {\hat \lambda}_k  / k $ where ${\hat \lambda}_k$ is the fixed point of the above function and observe that $\Psi (\cdot)$ represents similar function, but considering exact blocking probability  \eqref{Eqn_PB}.  For further analysis, we relax $k$ to be a real value between $(N/2, N)$. We immediately have the following result:
	\begin{lemma}
\textit{i)	 There exists  a  ${\bar \Lambda}$ such that, 
the function 	 $\psi$ is increasing with $k$, for any $\Lambda \ge {\bar \Lambda}$. \\
ii) For any  $\Lambda \ge {\bar \Lambda}$,  under second order approximation, the unique maximizer in \eqref{Eqn_kstar} is given by   $k^*  = \sum_{i=1}^{n-1} N_i $.\\
iii) For all such  $\Lambda$, the  partitions $\P = \{C,  {\cal N}\backslash C \}$,  with $  N/2 \le N_{C} \le k^*$ are the only  2-partitions that are stable, under second order approximation.}
\label{heavy_traffic}
	   \end{lemma}
	   \textbf{Proof: }See Appendix C. \eop

{\bf Accuracy of the approximation}   
We now prove that the above result is also true without approximation using \textit{maximum theorem} \cite{maximum}. We will show that both the fixed points converge towards each other and that there exists  a  ${\bar \Lambda}$ such that,  stable partition considering true blocking probability equals that derived with second order approximation.

%\subsubsection{Equivalence of First Order Heavy Traffic Erlang B approximation and the exact Erlang B}
 Consider   $y \in [\epsilon, 1-\epsilon]$ for some small $\epsilon >0$ and define the following function: 
	
	\vspace{-4mm}
	{\small\begin{equation*}
	g(y, \theta ) = \left \{  \begin{array}{llll}
	 \sqrt{\frac{1}{\theta} } \Bigg(\sum_{j=0}^{k-1} \frac{( y/\theta)^{j-k}}{j!}k! \\ -  \sum_{j=0}^{(N-k)-1} \frac{ [(1-y)/\theta]^{j-(N-k)}}{j!}(N-k)! \Bigg)^2,  \hspace{-2mm}&\mbox{if } \theta  > 0 \\
	 (y^{-1}k-(1-y)^{-1}(N-k))^2  &\mbox{if } \theta = 0. 
	 \end{array}  \right . 
	\end{equation*}}
	Observe from \eqref{Eqn_PB}  that   $y = \frac{\lambda_1}{\Lambda}$ (the normalized WE)  is the unique zero of the  function  $g(.)$ when $\Lambda = 1/\theta$, uniqueness given by Theorem \ref{Thm_WE}, 
	and the solution by first order approximation is zero of $g$ when $\theta = 0$.  
	It is clear that $g$ is a jointly continuous mapping\footnote{If $\epsilon$ is such that the true WE does not fall in interval $[\epsilon, 1-\epsilon]$, 
 then we would have some other points as the minimizers of $g(.)$ (by continuity and compactness), but eventually (with large enough $\Lambda$) we will have unique zero of $g(.)$ which is derived from unique WE of the original problem.} over $[\epsilon, 1-\epsilon] \times [0, B]$ (for any $B < \infty$). Define,
	\begin{align*}
	g^{*}(\theta) & \triangleq  \max_{y \in [\epsilon,1-\epsilon]} g(y,\theta) \mbox{ and }
	y^{*}(\theta) & \triangleq \arg\max_{y \in [\epsilon,1-\epsilon]} g(y,\theta).
	\end{align*}
	Then, by Maximum Theorem, $y^{*}(1/\Lambda) \to y^{*}(0)$ as $\Lambda \to \infty$. In other words we have:
	\begin{equation}
 \frac{\lambda_1 (\Lambda)}{\Lambda} \to \frac{k}{N} \mbox{, or equivalently, }  \frac{1}{\Lambda} \Bigg|\frac{\lambda_1^*}{k}-\frac{\Lambda}{N}\Bigg| \to 0
 \label{Eqn_max_proof}
	\end{equation}
	Using exactly similar logic, one can show that the WE using  second order approximation also converges towards that of the first order approximation, and hence that  the differences between the  WE obtained using second order approximation and that   obtained using true blocking probability \eqref{Eqn_PB}  converge towards each other.  Using this we prove:
	
%	By Theorem \ref{Thm_WE}.(iii), for any $\Lambda$,   $ \lambda_k   / \Lambda >  k/ N$. 
%	{\color{red}We need to show that zero of the WE lies in the interval $[\epsilon,1-\epsilon]$}
	
}{
We use the well known Maximum theorem (\cite{maximum}) and second order approximation to derive the following result. 
}

\begin{theorem} 
\label{Thm_approx}
 \textit{There exists  a  ${\bar \Lambda}$ such that,   only     $k^*  := \sum_{i=1}^{n-1} N_i $  optimizes   \eqref{Eqn_kstar}  for our queueing system with any $\Lambda \ge {\bar \Lambda}$. 
   Further the only partitions that are  part of a stable configuration (under RB-PA/RB-IA)  are the 2-partitions $\P = \{C,  {\cal N}\backslash C \}$,  with $  N/2 \le N_{C} \le k^*$.}
\end{theorem}
\Proofs{
{\bf Proof:} 
By the above arguments for any 2-partition  $\P = (C_1, C_2)$ with $N_{C_1}=k$, 
the 
 the WE obtained using second order approximation \eqref{Eqn_fixed_point_eq}, represented by   $ {\hat \lambda}_k$, and the one obtained  using the exact blocking probability  \eqref{Eqn_PB} converge towards each other for all $k$ as $\Lambda \to \infty.$
 Consider ${\bar \Lambda} $  further large in Lemma \ref{heavy_traffic} such that  (possible by finiteness)
$$\frac{1}{\Lambda}| {\hat \lambda}_k (\Lambda) - \lambda_k (\Lambda) | < \delta \mbox{ for all  possible } k, \mbox{ and for }   \Lambda \ge {\bar \Lambda}, $$ 
where $\delta >0$ is sufficiently small so that the conclusions of the theorem follow from that in Lemma \ref{heavy_traffic}  (possible by finitely many values of $k$), after observing that monotonicity of 
$\Psi$ with respect to $k$ for any given $\Lambda$ is equivalent to monotonicity of $\Psi / \Lambda$ with respect to $k$. \eop 
}{
{\bf Proof} is provided in technical report \cite{TR}. 
\eop}

\subsection{Light Traffic}
We now consider light traffic regimes, and derive the following result
using similar logic as before:

\begin{lemma}
  \textit{There exists a ${\underline \Lambda} >0$, such that for all
    $\lambda < {\underline \Lambda}$, we have: i)
    ${\underline k}^* := \min \{ N_C : N_C > N/2, C \subset {\cal N}
    \}$ optimizes \eqref{Eqn_kstar} for our queueing system; and ii)
     Further the only partitions that are  part of a stable configuration (under RB-PA/RB-IA)  are the 2-partitions $\P = \{C,  {\cal N}\backslash C \}$,  with $  N/2 \le N_{C} \le k^*$.}
  % \textit{There exists a $\bar{\lambda}_1(\epsilon)$ such that for
  % all $\lambda_1 \leq \bar{\lambda}_1(\epsilon)$, partition with two
  % coalitions with one of the coalitions having
  % $\lfloor \frac{n}{2} \rfloor +1$ agents is the coalitionally
  % stable partition.}
  \label{low_traffic}
\end{lemma}	
\textbf{Proof: } See Appendix C. \eop

Thus in both (light as well as the heavy) the traffic regimes, stable
partitions are the 2-partitions with $N/2 \le k \le k^*$.  Further
observe that assumption {\bf A}.1 is satisfied for both these
regimes. Hence the random dynamics under RB-IA, in either case, converges and
stops at one of such 2-partitions.

\Cmnt{
	\subsection{QED-Regime}	
	Assuming $\mu=1$, we define $\rho = \Lambda/N$. In this regime, we aim to find stable partitions when $\rho$ is a constant.
	%$\Lambda = \rho*N$
For a constant $\rho$ and a specific value of $\Lambda$ and $N$, the game is similar to the previous one. Using Theorem \ref{Thm_duo_mono} we know partitions of size more than 2 are not stable.
	
	Considering any 2-partition, we define $\gamma$ to be the fraction of servers in $C_1$ such that
	$k = \gamma * N$ and $N-k = (1-\gamma)N$.

	$$
	B^{-1}(\rho, N) = e^{\rho}\rho^{-N} \int_{\rho}^{\infty} t^N e^{-t} dt
	$$
	
	$$
	B^{-1}(\rho N, N) = e^{\rho N}{\rho N}^{-N} \int_{\rho N}^{\infty} t^N e^{-t} dt
	$$
	
	\begin{eqnarray}
	B^{-1}(\rho N, \gamma N) &=& e^{\rho N}{\rho N}^{-\gamma N} \int_{\rho N}^{\infty} t^{\gamma N} e^{-t} dt \nonumber \\
	& = & \int_{\rho N}^{\infty} \Big(\frac{t}{\rho N}\Big)^{\gamma N} e^{-t + \rho N} dt \nonumber 
	\end{eqnarray}
	Let $\frac{t}{\rho N} = z$. Then, $\frac{dt}{\rho N} = dz$.
	\begin{eqnarray}
	B^{-1}(\rho N, \gamma N) &=& \int_{1}^{\infty} z^{\gamma N} e^{-(z-1)\rho N} \rho N dz \nonumber 
	\end{eqnarray}
	Let $z-1 =y$. Then $dz =dy$.
	\begin{eqnarray}
	B^{-1}(\rho N, \gamma N) &=& \int_{0}^{\infty} (y+1)^{\gamma N} e^{-y\rho N} \rho N dy \nonumber \\
	= E[(Y+1)^{\gamma N}] \nonumber
	\end{eqnarray}
	if $Y = exp(\rho N)$ or $\rho N Y = exp(1)$.
	
	After some calculations, in the limits we have:
	$$
	B = 1-\frac{\gamma}{\rho}
	$$
	Thus, for any 2-partition at WE, we require:
	\begin{eqnarray}
	\frac{\gamma}{\rho_1} &=& \frac{1-\gamma}{1-\rho_1} \nonumber \\
	\rho_1&=&\gamma \nonumber
	\end{eqnarray}
	Thus in the limits, blocking probability is 0. }

\Cmnt{\subsection{Schedulers for Partition Form Games}
\begin{itemize}
	\item $\mathcal{N}$ is any coalition and $\mathcal{P}$ is any partition.
	%\item $\boxed{ \phi_i^{\mathcal{N}, \mathcal{P}} \, \forall \, i \in \mathcal{N},\, \forall \, \mathcal{N} \in \mathcal{P} \text{ and } \forall \, \mathcal{P} }$ is the scheduler, i.e., division of worth among members of $\cal{N}$ in partition $\mathcal{P}$.
	\item $\bar{\lambda}_C$ and $\underline{\lambda}_C$ is the maximum and minimum utility respectively that coalition $C$ can obtain irrespective of the arrangement of agents outside this coalition
	\item An allocation rule is defined as follows:
	\begin{equation}
	\{\{\phi_i^{\mathcal{N}, \mathcal{P}}\}_{i \in \mathcal{N}} ; \, \forall \, \mathcal{N} \in \mathcal{P} \text{ and } \, \forall \, \mathcal{P}\} 
	\end{equation}
	 is the scheduler, i.e., division of worth among members of $\cal{N}$ in partition $\mathcal{P}$.
	We know by definition:
	\begin{equation}
	\sum_{i \in \mathcal{N}} \phi_i^{\mathcal{N},\mathcal{P}} = {\lambda}^{\mathcal{N}, \mathcal{P}}
	\end{equation}
\end{itemize}

\subsection{Fair Sharing Scheme}
We say a scheduling scheme is fair if it satisfies - ``Whenever a merger gives better utility the division among the members of the merger coalition should become strictly better than before for all the members of the merger." To be more precise,

\noindent
If there exists a disjoint collection of coalitions $C_1,C_2, \cdots,C_n$ in any partition  $\mathcal{P}$  with merger coalition  $\mathcal{N} := C_1\cup C_2 \cup \cdots \cup C_n$ having strictly better utility ``even against the worst leftover partition $\mathcal{P}'$", 
\begin{equation}
\sum_{i \in \mathcal{N}} \phi_i^{\mathcal{N},\mathcal{P}'}  > \sum_{i \in \mathcal{N}} \phi_i^{C^i,\mathcal{P}} 
\label{condition}
\end{equation}
where $\mathcal{P}'$ is the partition containing $\mathcal{N}$ and the remaining agents arranged to hurt the coalition $\mathcal{N}$ most and $C^i$ is the coalition containing agent $i$ in partition $\mathcal{P}$.

\noindent
Then, the scheduler should allocate such that each member of the merger strictly improves, i.e., 
\begin{equation}
\phi_i^{\mathcal{N},\mathcal{P}'}  >  \phi_i^{C^i,\mathcal{P}} \, \, \forall \, i \in \mathcal{N}
\label{fair}
\end{equation} \Cmnt{$\in \mathcal{P} C \subset \cal{S}$ and there exists $S_1, S_2, \cdots, S_n$ (disjoint subsets in $\mathcal{S} / C$) in partition $\mathcal{P}$ such that if
	\begin{equation}
	\underline{\lambda}_{C\cup S_1, \dots, \cup S_n} > \bar{\lambda}_C+\sum_{S_i} \bar{\lambda}_{S_i} \text{ where } C \cup S_1, \dots, \cup S_n = \cal{N}
	\label{condition}
	\end{equation}
	then for all $C' \subset \cal{N}$, we have
	\begin{equation}
	\sum_{i \in C'} \phi_i^{\cal{N}} > \bar{\lambda}_{C'}
	\end{equation}}

\subsubsection*{Examples}
\begin{itemize}
	\item Proportional.
	\item Extra utility is divided equally or in such a way that each player gets strictly better. 
	\item Can we say it lies in core (extended for Partition form games?)
	%\item If 
\end{itemize}
\begin{lemma}
	Partitions with cardinality greater than 2 are not coalitionally stable partitions under fair sharing scheme.
\end{lemma}
\noindent
\textbf{Proof: }Consider a partition  $\mathcal{P} = \{C_1,C_2,\cdots, C_n\}$ with cardinality greater than 2. Let $\mathcal{C}$ be the coalition containing all coalitions of $\mathcal{P}$ except one. Then equation \eqref{condition} is satisfied for such a $\cal{C}$ because of the following argument:

\noindent
Each of the coalition before merging had same blocking probability at WE. When they merge, the blocking probability of the new coalition reduces which in turn increases the arrival rate to this coalition at WE.
\begin{equation*}
\sum_{i \in \mathcal{N}} \phi_{i,\mathcal{C}}^{\mathcal{P}'} = \underline{\lambda}_{\mathcal{C}}^{\mathcal{P}'}  > \sum_{C_i } \lambda_{C_i}^ \mathcal{P} = \sum_{i \in \mathcal{C}} \phi_{i,C^i}^{\mathcal{P}}
\end{equation*}

\noindent
Then, each player in the new coalition gets better utility by equation \eqref{fair} under any fair scheduler and hence such a partition is blocked by $\mathcal{C}$ and hence not coalitionally stable.
\eop}

\Cmnt{\noindent
The \textbf{$\alpha$-core} of this game is the set of all coalitions such that no coalition can $\alpha$-block it.

\subsection{Duopoly in $\alpha$-core}
\begin{itemize}
	\item Using same arguments as in Lemma \ref{atmost2}, coalitions of a partition with size more than 2 are $\alpha$-blocked.
	\item When all players merge, it is not possible for each player to get strictly better utility since the total utility is still the same and hence it is also $\alpha$-blocked. \textbf{Constant-sum game}
\end{itemize}

\section{Shapley Value}

\textit{Parameters:}

\begin{enumerate}
	\item $\Lambda$ =25
	\item $N_1$ =8, $N_2 = 4$, $N_3$ = 3 and $N_4$=2
	\item $N= 50000$
\end{enumerate}

\textit{Results:}
\begin{enumerate}
	\item When partition $\{\{1,3\},\{2\}, \{4\}\}$
	utilities of player 1, 2, 3 and 4 are 13.1161882, 5.4227001, 4.2924711 and 2.1686405.
	\item When partition $\{\{1,3,4\},\{2\}\}$ utilities of player 1, 2, 3 and 4 are 12.8511764, 5.1138910, 4.3390555 and 2.6958771 .
\end{enumerate}
We can observe that the utilities obtained by coalition with players 1,3 and 4 in the second partition is more. However, player 1 obtains smaller share which contradicts our definition of fair schedulers.

\textit{Intuition:}}

\section{Numerical Computations}
		
In this section, we simulate two systems with $5$ agents. The agents in first and second system have 9, 7, 6, 5, 3 and 10, 7, 6, 5, 4 servers respectively. From left sub-figure of  Figure \ref{fig:optimal_k}, we have that $k^*$  is a monotonically increasing function of $\Lambda$ in both cases. The corresponding right sub-figure shows blocking probability as a function of $\Lambda$ in both cases.
 \begin{figure}[h]
 %\centering
 \begin{minipage}{.5\textwidth}
   %\centering 
    \includegraphics[width=.4\linewidth]{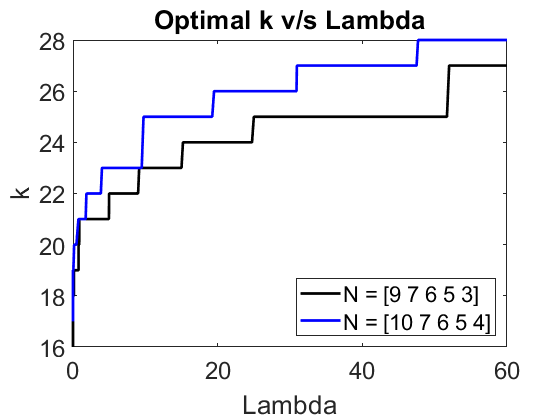}
   %\caption{Optimal $k^*$ v/s $\Lambda$}
  
 \end{minipage}%
  \begin{minipage}{.5\textwidth}
   %\centering
   \hspace{-40mm}
 \includegraphics[width=.4\linewidth]{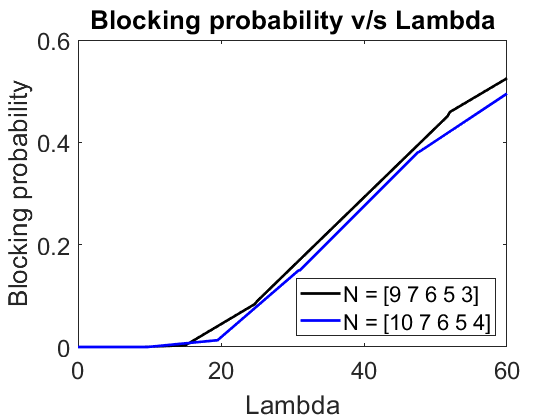}
    %\caption{$B$ v/s $\Lambda$}

 \end{minipage}
 \caption{Optimal $k^*$ v/s $\Lambda$ (left) and $B$ v/s $\Lambda$ (right)  \label{fig:optimal_k}  \label{fig:blocking_prob}}
 \end{figure}
 One may observe that the end points of the left sub-figure, i.e., under heavy and low traffic matches with the ones derived in Theorem \ref{Thm_approx} and Lemma \ref{low_traffic}.

\section{Conclusions}

We consider a queueing system with several strategic service providers with different server capabilities.  They are on lookout for collaboration opportunities  that improve their individual payoffs. 
The customer base responds to any operational arrangement  formed by such collaborations, the   customer arrivals are  split across various operational units according to the well known Wardrop equilibrium that equalizes the steady state blocking probability of all the units. Any  operational configuration is challenged by new coalition formed by mergers or splits, and the former is dissolved if the new coalition finds benefit. A configuration is stable if there is no coalition to challenge it. We defined three notions of stability and our major findings are: a)  configurations with more than two coalitions  are never stable; b) grand coalition can be stable depending upon the payoff allocations and the notion of stability, only  if there exists a single dominant player with more than half the server capacity of the system; and c) some configurations with two coalitions are stable, depending upon the notion of stability and  the payoff allocations.  We also consider an initial model with dynamic coalition formations and showed the convergence of the same under one notion of stability.   This work  just opened an array of questions that need exploration.

\section*{Appendix A:  Characteristic Form games }
\Proofs{
Both 
partition-form and classic (non-partition) cooperative games can be described in }{A game in}
   characteristic form \cite{aumann1961} \Proofs{}{ can be defined } using  the tuple,  $(\mathcal{N},  \nu,\mathcal{H})$, 
where: 
a)
$\nu$ is called a characteristic function and  for any $C \subseteq \mathcal{N}$,  
$\nu(C)$  denotes the set  of all possible
 payoff vectors of dimension $n$ that agents in  $C$ can jointly achieve; 
b)
$\mathcal{N}$ denotes the set of $n$ agents; 
 and
 c) 
$\mathcal{H}$ is the set of all possible payoff vectors of dimension $n$ (such vectors are also referred to as allocation vectors in literature), which are achievable.  

%All the pay-off vectors are of dimension $n$ unless stated otherwise.
In this appendix, we provide the details of how our problem can be recast as a characteristic game.  

%Mathematically, one can define
%\begin{equation}
%\nu(C) = \cup_{\mathcal{P}: C \in \mathcal{P}} \tilde{\nu}(\mathcal{P})
%\end{equation}
%where

Let  ${\cal F} (\P) $ be the set of all feasible payoff vectors %of dimension $n$ 
under partition $\P$,  these are the vectors that satisfy the following:     the sum of payoffs of all  agents in any coalition $S$ is less than or equal to that obtained by $S$ under partition $\mathcal{P}$ at WE,  $\lambda_S^\mathcal{P}$. Thus   %\vspace{-5mm}
\begin{equation}
{\cal F} (\P)   := \left \{\x = [\xs_i] : \sum_{i \in S} \xs_i \leq \lambda_S^\mathcal{P} \, \forall \, S \in \P   \right \}.
\end{equation}
Thus $\mathcal{H}$, the set of all achievable/feasible payoff vectors is,
\begin{equation}
\mathcal{H} = \cup_{\mathcal{P}} {\cal F} (\P) .
\end{equation}

{\bf  Characteristic function using pessimistic rule:}
To study the stability aspects, one needs to understand if a certain coalition can `block' any payoff vector.
Blocking by a coalition implies that coalition is working as an independent unit and has an anticipation of the value it can achieve (irrespective of arrangements of others). 
If the division of this anticipated value among the members of the coalition, under any given allocation rule,  renders the members to achieve more than that in the given payoff vector, then the coalition has tendency to oppose the  payoff vector.

The  characteristic function precisely describes the set of all possible divisions of the anticipated worth  of any coalition.

There are many anticipatory rules to define characteristic function for partition form games. 
The above described rule is the well known   \textit{pessimistic anticipation rule} \cite{pessimistic}, where  the agents in deviating coalition $C$ assume that the outside agents arrange themselves to hurt the agents in $C$ the most. 

Towards specifying such a characteristic function,  first 
 observe that  the minimum utility that coalition $C$ can achieve irrespective of the arrangement of the agents outside this coalition is given  by:  %\vspace{-5mm}
\begin{equation}
%\hspace{16mm}
{\underline \nu}_C   := \min_{ \x \in {\cal F} (\P) : C \in \P }  \sum_{i \in C}  \xs_i  .
\label{Eqn_pec_bnd} 
\end{equation}
With this definition,  $\nu(C)$, the set of possible payoff vectors that agents in $C$ can jointly achieve independent of the arrangement of outside agents is given by,

%\vspace{-5mm}
{\small \begin{equation}
\nu(C) = \left \{\x \in \mathcal{H}: \sum_{i \in C}  \x_i \leq \underline{\nu}_C \right \}.
\label{Eqn_pec_nu}
\end{equation}}
%where $\underline{\nu}_C$ is the minimum utility that coalition $C$ can achieve irrespective of the partition, i.e., 
%\begin{equation}
%\underline{\nu}_C = \min_{\mathcal{P}: C \in \mathcal{P}} \nu(C, \mathcal{P}) \, \forall \, C \in \mathcal{P}
%\end{equation}

\Proofs{
With these definitions in place, we can now define when  a payoff vector is \textit{blocked by a coalition}.}{}

\textit{Blocking:} 
A payoff vector $\x \in \mathcal{H}$ is blocked by a coalition $C$ if there exist a payoff vector $\textbf{y} \in \nu(C)$ such that
\begin{equation*}
y_i > \xs_i \, \, \forall \, i \in C.
\end{equation*}

Next, we define \textit{$\alpha$-core}, which is an extension of the classical definition of \textit{core}, for transferable utility games (in non-partition form games).

\textit{$\alpha$-core: } It is the set of all feasible payoff vectors, i.e., $\x \in \mathcal{H}$ such that it is not blocked by any coalition $C$. 
In other words   $\alpha$-core is coalitionally rational:   it consists of all feasible payoff vectors such that no coalition of agents can deviate and achieve better.

\Proofs{

\onecolumn

\section*{Appendix B:  Proof of Theorem \ref{Thm_WE}}

\textbf{ Proof of  Existence and Uniqueness:}
Let the size of a partition be denoted by $p$.
The first step of this proof is to show the existence and uniqueness of WE for the case when $p=2$.  In the next step,  using induction we prove the existence for any general $p=m>2$ using the results for $m-1$. In the third step we   show the continuity of the WE, to be precise the arrival rates at WE for $m$. The last step attributes to the uniqueness of our solution.

\textit{Step 1: Existence and Uniqueness of WE for $p=2$} 

%We begin with 2-partition, i.e., $p=2$. 
To obtain WE, the following equation need to be solved:
$$
B_{C_1}(N_{C_1},a_{C_1}) = B_{C_2}(N_{C_2},a_{C_2}).
$$
Define a function $f:= B_{C_1}(N_{C_1},a_{C_1}) - B_{C_2}(N_{C_2},a_{C_2})$. Then, $f$ is a function of $\lambda_{C_1}^\mathcal{P} \in [0,\Lambda]$ since $\lambda_{C_2}^\mathcal{P}= \Lambda-\lambda_{C_1}^\mathcal{P}$. 
\begin{itemize}
	\item 
	At $\lambda_{C_1}^\mathcal{P}=0$ we have $B_{C_1}(N_{C_1},a_{C_1})=0$ and $B_{C_2}(N_{C_2},a_{C_2}) >0$, thus $f(0) <0$.
	
	\item At $\lambda_{C_1}^\mathcal{P}=\Lambda$ we have $B_{C_1}(N_{C_1},a_{C_1})>0$ and $B_{C_2}(N_{C_2},a_{C_2}) =0$, thus $f(\Lambda) >0$. 
\end{itemize}
Then, $B_{C_1}(N_{C_1},a_{C_1})$ and $B_{C_2}(N_{C_2},a_{C_2})$ are polynomial functions with denominator $> 1$ and hence are continuous functions. This implies that $f$ is a continuous function.

Thus, $f$ satisfies the hypothesis of Intermediate Value Theorem (IVT). Using IVT, there exists a value of $\lambda_{C_1}^\mathcal{P} = \lambda^* \in (0,\Lambda)$ such that $f(\lambda^*)=0$.
The uniqueness of $\lambda^*$ follows since $B_{C_1}(N_{C_1},a_{C_1})$ and $B_{C_2}(N_{C_2},a_{C_2})$ are strict increasing functions of $\lambda_{C_1}^\mathcal{P}$ and $\lambda_{C_2}^\mathcal{P}$ respectively. 

 \textit{Step 2: Existence for general $p = m > 2$}
 
 To prove the existence for any general $m>2$, we assume that a unique WE exists for $p=m-1$, i.e., $\lambda_{C_1}^{\P}, \cdots, \lambda_{C_{m-1}}^{\P}$ with corresponding common blocking probability $B^*$.
 
 With $m$ units we can initially fix $\lambda_{C_{m}}^{\P} = 0$ and obtain WE corresponding to the remaining units, which we have assumed to exist. With increase in $\lambda_{C_m}^{\P}$, $\Lambda - \lambda_{C_m}^{\P}$ which is the total share of remaining agents, decreases. From part i) of this theorem, we know that the corresponding WE solution for these agents also decreases. This implies that the common blocking probability for $C_1, \cdots, C_{m-1}$ reduces while blocking probability of $C_m$ increases. Using similar arguments as above and treating $C_1, \cdots, C_{m-1}$ as one while defining function for IVT, one can show that WE exists.

\Cmnt{Next we use induction to show that a split among $(m+1)$ units exists at WE if we have a split among $m$- units. 
The next case is when all the three agents do not pool their resources such that the WE equation is:
$$
pb_1 = pb_2 = pb_3
$$
$$
	pb_i = \frac{\frac{a_i^{N_i}}{N_i!}}{\sum_{j=0}^{N_i} \frac{a_i^{j}}{j!}}
	$$ 
	At equilibrium, we have
	$$
	\frac{\frac{a_1^{N_1}}{N_1!}}{\sum_{j=0}^{N_1} \frac{a_1^{j}}{j!}} =  \frac{\frac{a_2^{N_2}}{N_2!}}{\sum_{j=0}^{N_2} \frac{a_2^{j}}{j!}} =  \frac{\frac{(a-a_1-a_2)^{N_3}}{N_3!}}{\sum_{j=0}^{N_3} \frac{(a-a_1-a_2)^{j}}{j!}}
	$$
	We can say it as $pb_1 = pb_2 = pb_3$.
	\subsubsection{Existence and Uniqueness of Wardrop Equilibrium}
	
	\textbf{Proof: }
	At Wardrop Equilibrium, we require
	$$
	pb_{1} = pb_2 = pb_3
	$$
Since the WE exists for a split among $n$ units, we know that for a fixed value of $\lambda_{C_1}^\mathcal{P}$ there exist $\lambda_{C_2}^\mathcal{P},\cdots,\lambda_{C_{m+1}}^\mathcal{P}$ such that the blocking probability of each unit in partition, i.e. $B^*$ is same (since $\Lambda-\lambda_{C_1}^\mathcal{P}$ is again a known constant). So we begin with $\lambda_{C_1}^\mathcal{P}=0$ and obtain WE solution. Now as we increase $\lambda_{C_1}^\mathcal{P}$, total arrival rate to be split ($\Lambda-\lambda_{C_1}^\mathcal{P}$) amongst units $C_2, \cdots, C_{m+1}$ decreases which in turn reduces their individual splits ({\color{red}From Lemma \ref{WE_2waysplit_monotonic}}). 

Now since blocking probability is an increasing function of arrival rates, $B_{C_1}(N_{C_1},a_{C_1})$ increases with increase in $\lambda_{C_1}^\mathcal{P}$, while $B^*$ reduces. Thus, we have a similar case as above (with partition size 2) where with increase in $\lambda_{C_1}^\mathcal{P}$ one function is increasing while the other is decreasing.
Hence by the similar arguments as above (which includes that $B_{C_1}(N_{C_1},a_{C_1}) \uparrow 1$ and $B_{C_2}(N_{C_2},a_{C_2}) \downarrow 0$  as $\lambda_{C_1}^\mathcal{P} \uparrow \Lambda$) we know there exists a unique WE (uniqueness follows from strict monotonic property of Erlang-B).We know the solution to two-way WE split exists and is unique. Assuming it is true for $n$-way WE split. We want to show that it is true for $(n+1)$-way split as well.

\noindent 
Suppose the arrival rate for the new coalition is $0$. Then, since $n$-ways split exists and is unique we have a solution. Now we increase arrival rate to the new coalition which increases its blocking probability. While the amount of total arrival rate left to split is reduced, which reduces their splits by Lemma \ref{WE_2waysplit_monotonic}; and hence their blocking probability is reduced. Using the Intermediate Value theorem and the fact that the function is strictly monoonic we have the existence and uniqueness of the solution.}

\textit{Step 3: Continuity of Optimizers, i.e., WE:} 
Consider the following function  for $m$ units in partition $\P$: 
\begin{equation}
g(\Lambda,\lambda) := \sum_{C_j \in \mathcal{P}; 1<j\leq m} (B_{C_1}-B_{C_j})^2,
\end{equation}
where $\lambda$ is the vector of arrival rates for all $C_j \in \P$. Then, we define
$g^*(\Lambda,\lambda^*) = \min_{\{\lambda: \sum_j \lambda_j = \Lambda \} }g(\Lambda,\lambda) $.
Observe that the (unique) minimizer $\lambda^*$ of the function $g$ is the (unique) WE for our queueing model, and that the function $g$ is jointly continuous. Thus, using Maximum Theorem we have that $g^*$ and $\lambda^*$ is continuous in $\Lambda$.

\textit{Step 4: Uniqueness of WE }
To prove the uniqueness of the WE, we assume the contradiction. Accordingly, we can have the following cases: 

 \textit{Case 1: There exist multiple WEs with same common blocking probability $B^*$} 
 
  This implies that some of the units in partition are obtaining different arrival rates in the multiple WEs such that they have common $B^*$. However, this is not possible since blocking probability is an increasing function of arrival rates. Thus, the unit with higher arrival rate in one of the WEs should have higher blocking probability.
 
 \textit{Case 2: There exist multiple WEs with different common blocking probability $B^*$ and ${\hat B}^*$} 
 
 Without loss of generality, we can assume that $B^* < \hat{B}^*$. This implies that the arrival rates to the units with common blocking probability $\hat{B}^*$ is more (since blocking probability is an increasing function of arrival rate). However, the total arrival rate is fixed at $\Lambda$ which implies that one of the WE does not satisfy $\sum_{C_j \in \P} \lambda_{C_j}^{\P} = \Lambda$.
\eop

%\begin{lemma}
%\textit{Let $\P = \{C_1, \cdots, C_k\}$ be any partition.  i) If $\P'  $ is formed by merger of two coaltions  $C_j, C_l$ of $%%\P$, with  the rest remaining  intact, then 
%$$
%\lambda_{C_l}^{\P} + \lambda_{C_j}^{\P} < \lambda_{C_l \cup C_j}^{\P'}.
%$$ 

%\label{Lem_merger_split_arrival}
%\end{lemma}

%This result follows from the fact that merger/split of any two coalitions lead to reduction/increase in overall blocking probability. This in turn implies that the arrival rates increase /decrease with a merger/split, since the blocking probability is a monotonically increasing function of arrival rates. \eop

\textbf{Proof of  All units used}  %Let $\lambda_{C_1}^\mathcal{P}, \cdots, \lambda_{C_k}^\mathcal{P}$ be the individual arrival rates corresponding to partition $\mathcal{P}$ at WE for the coalitions $C_1, \cdots, C_k$ respectively with the total arrival rate $\Lambda>0$.
For contradiction, let us assume that the customers split themselves amongst some strict subset of units of partition $\mathcal{P}$. Then, each unit with zero arrivals have a zero blocking probability while units with non-zero arrivals have some strict positive blocking probability. However, this contradicts the fact that the coalitions having zero arrivals should have a higher blocking probability than others at WE. \Cmnt{This implies if any arrival happened to any unused unit its blocking probability will be strictly smaller than the other used units, which contradicts the first principle of WE.} %Then, atleast one of these 
%units obtain higher arrival rate than their counterparts (when the customers split amongst all units of $\mathcal{P}$). The increase in arrival rate leads to an increase in its blocking probability. Using similar arguments as in {\color{red} part of this Lemma } we can show that each unit with non-zero arrivals has higher arrival rate than before and hence a higher common blocking probability which is a contradiction to the system optimality of WE.

Hence at WE, each of the units in partition $\mathcal{P}$ obtain non-zero arrival rates. \eop

\textbf{Proof of part (i)}  Let $\lambda_{C_1}^\mathcal{P}, \cdots, \lambda_{C_k}^\mathcal{P}$ be the individual arrival rates corresponding to partition $\mathcal{P}$ at WE  (satisfies \eqref{Eqn_WE_properties}) for the coalitions $C_1, \cdots, C_k$ respectively with the total arrival rate $\Lambda>0$. Let the   corresponding  common  blocking probability be  $B^*$. When the total arrival rate is increased to $\Lambda'$, the individual arrival rates to the providers at WE are changed to $\lambda_{C_1}^{'\mathcal{P}},\cdots, \lambda_{C_k}^{'\mathcal{P}}$ and the   corresponding  common  blocking probability is changed to  $\hat{B}^*$. Note that these splits to the individual operating units must satisfy:
\begin{equation}
\sum_{i=1}^k \lambda_{C_i}^{\mathcal{P}} = \Lambda \text{ for any partition } \mathcal{P}.
\label{total_conserved}
\end{equation} 

\noindent
Next we will show that $\lambda_{C_j}^{'\mathcal{P}} \leq \lambda_{C_j}^{\mathcal{P}}$ is not possible for any  $C_j \in \mathcal{P}$. Using equation \eqref{total_conserved}, we know that atleast one of the units have higher individual arrival rates at new WE, i.e,
\begin{equation}
\lambda_{C_j}^{'\mathcal{P}} > \lambda_{C_j}^{\mathcal{P}} \text{ for atleast one } C_j \in \mathcal{P}.
\label{monotonic}
\end{equation}
This means that the common blocking probability at new WE is increased, i.e., $\hat{B}^* > B^*$. Now since blocking probability is a strictly increasing function of arrival rates, we have that arrival rate to each coalition is increased at new WE for $\Lambda'$, i.e., $\lambda_{C_j}^{'\P} > \lambda_{C_j}^{\P}$ for all $C_j \in \P$. 

%This implies that the blocking probability  of all such providers is increased. Thus, we can have the following scenarios:
%\begin{enumerate}
	%\item All providers follow equation \eqref{monotonic}. This proves the lemma.
%	\item Some of the remaining providers get same individual arrival rate as before, i.e., $\lambda_{C_i}^{'\mathcal{P}} = \lambda_{C_i}^{\mathcal{P}}$ at WE. This implies its blocking probability remains same. %However, atleast one of the providers following equation \eqref{monotonic} has higher blocking probability than agent $i$. Thus, such a split cannot emerge at WE and hence a contradiction.
	%\item Similarly, if some of the remaining providers get lower individual arrival rate as before, i.e., $\lambda_{C_i}^{'\mathcal{P}} < \lambda_{C_i}^{\mathcal{P}}$ at WE. This implies its blocking probability is further reduced. %
%\end{enumerate}
%However, atleast one of the providers following equation \eqref{monotonic} has higher blocking probability than agent $i$. Thus, such a split cannot emerge at WE and thus a contradiction.

Hence, WE is an increasing function of  $\Lambda$. \eop

\textbf{Proof of part (ii)} Let $\lambda_{C_1}^\mathcal{P}, \cdots, \lambda_{C_k}^\mathcal{P}$ be the individual arrival rates corresponding to partition $\mathcal{P}$ at WE for the coalitions $C_1, \cdots, C_k$ respectively. %Consider merger of $C_1$ and $C_2$, i.e., $M = C_1 \cup C_2$. 
 Let the   corresponding  common  blocking probability be  $B^*$.  Observe that the blocking probability of $C_i$ and $C_j$ units also equals  $B^*$, and hence    the merger $M = C_i \cup C_j \neq \mathcal{N}$ has strictly smaller  blocking probability, i.e.,  $B_M < B^*$,  if the joint arrival rate was $\lambda_{C_i}^\P + \lambda_{C_j}^\P$. 
From \eqref{Eqn_PB} the  blocking probability is a strictly decreasing function of arrival rate. Thus the new WE after merger is formed with a (strict) bigger arrival rate to the merger, as again at the new WE the new blocking probabilities of all coalitions $C \in {\P}'$ \Cmnt{$C \in \P\backslash \{C_i, C_j\}$ and $M$} should be equal by  \eqref{Eqn_WE_properties}\Cmnt{ Theorem \ref{Thm_WE}}.  \eop

\Cmnt{b) Similar arguments as a) can be given. With the only difference %being 
that on splitting of coalition $C$ into $S$ and $S'$ each of these split coalitions (i.e., $S$ and $S'$) has higher blocking probability and thus, strictly smaller arrival rates.} %and thus, arrival rate to them is decreased at new WE. 

\textbf{Proof of  part  (iii)} 
 Consider a system with identical servers. We know that when any number of identical servers combine with their arrival rates, the combined blocking probability reduces. This reduction is more when the number of servers combining are more, i.e., %i) Let us begin with the case where coalition with $N-k$ and $k$ servers gets exactly $(N-k)/N$ and $k/N$ share of total arrival rate $\Lambda$ at WE respectively. This implies that each server obtain same , i.e., $1/N$ share of $\Lambda$.
%
%We know that 
\begin{equation}
B(N,a) > B(LN,La) > B(MN,Ma).
\label{Eqn_common_result}
\end{equation}
where $0<L< M$ are constants, $B$ is the blocking probability, $N$ is the number of servers and $a$ is the offered load.
Now if we consider that the coalition with $N_{C_1}$ and $N_{C_2}$ servers gets exactly $N_{C_1}/N$ and $N_{C_2}/N$ share of total arrival rate $\Lambda$ at WE respectively.
Using equation \eqref{Eqn_common_result}, we have that coalition with $N_{C_1}$ servers has strictly smaller blocking probability. From \eqref{Eqn_WE_properties}, the blocking probability of each unit at WE is same. So, the arrival rate to coalition with $N_{C_1}$ and $N_{C_2}$ servers need to be increased and reduced respectively to achieve the WE. %This means that the arrivals to this coalitions need to be increased to obtain WE.
%For contradiction, let us assume that the  %This proof can be split into two parts:
	%
	%\begin{enumerate}
		%
		%\item

%Under proportional share based allocation rule, this implies that each server obtains $1/N$ share of $\Lambda$ at WE. Now consider merging of servers one by one to form both coalitions. We know that when any two servers are combined the overall blocking probability of this merged system reduced. However in larger coalitions we have more such mergers which implies that its blocking probability is smaller than the other coalition. Hence, such a split cannot be a WE. 

%The arrival rate to the larger coalition (i.e., with $k$ servers) needs to be increased since its blocking probability is smaller than the other coalition. 

Hence, coalition with $N_{C_1}$ and $N_{C_2}$ servers satisfy
$$
 \hspace{25mm} 
\frac{\lambda_{C_1}^{\P}}{N_{C_1}} > \frac{\Lambda}{N} > \frac{\lambda_{C_2}^{\P}}{N_{C_2}}.  \hspace{25mm} \mbox{ \eop}
$$		%Now coalition with $p$ agents has more servers with same  arrival rate as the other coalition.  Since, blocking probability reduces with more number of servers, the equal shares cannot be the WE.
	%	
		%\item \textit{When coalition with $N-k$ servers gets more than $(N-p)/N$ share of total arrival rate $\Lambda$ at WE.}
		%%With equal shares to both coalitions, we know that the smaller coalition has higher blocking probability. Now if we increase arrival rate to this coalition (which reduces the arrival rate to other coalition), this further increases the gap between the blocking probabilities of two coalitions. Hence, this also cannot be the WE.
	%\end{enumerate}
	%
%	But we know that the WE exists and hence the result is proved. 
%
\Cmnt{ii) Under PSA rule \eqref{Eqn_PSA},  agent $i$ at GC obtains $(N_i/N) \Lambda$ for any $i$.  However if some of the agents deviate together to form a coalition $C$ such that the following is true. 
\begin{equation}
N_C = \sum_{i \in C} N_i > \sum_{i \in \mathcal{N} \backslash C} N_i,
\end{equation}
then by part (i) of this lemma, $C$  obtains more than $(N_C / N)  \Lambda$, irrespective of the way ${\cal N}\backslash C$ agents operate; the worst utility is obtained by $C$ unit, when  all ${\cal N}\backslash C$  agents operate together as in part (i).   Thus under PSA rule \eqref{Eqn_GC_fair} is satisfied.  
Hence, PSA rule is UGC.}

\section*{Appendix C: Rest of the proofs}

\textbf{Proof of Theorem \ref{Thm_duo_mono}:}
Consider a partition  $\mathcal{P} = \{C_1,C_2,\cdots, C_k\}$ with cardinality greater than 2. Let $M$ be the merger coalition containing all coalitions of $\mathcal{P}$ except one, i.e.,
$$
M = \cup_{i=2}^k C_i \text{ and } \P' = \{C_1,M\}.
$$ Then from Theorem \ref{Thm_WE}.(ii)
\Cmnt{\noindent
Each of the coalition before merging has same blocking probability at WE. Using Lemma \ref{Lem_merger_split_arrival} repeatedly, %If two of these coalitions merge then the overall blocking probability of this merged coalition is reduced which in turn increases the arrival rate to this coalition at WE. This in turn reduces the arrival rate to $C_1$. We can continue such a merging stepwise until coalition $M$ is formed. At each step, 
we have that the arrival rate  obtained by $C_1$ in $\P'$ is smaller than in $\P$ and we have,
\begin{equation}
\lambda_{C_1}^{\P'} = \ulam_{C_1}^{\P'} <  \lambda_{C_1}^{\P}.
\end{equation}
This in turn implies that
\begin{equation}
\lambda_M^{\P'} > \sum_{C_i \in M} \lambda_{C_i}^{\P}.
\end{equation} 
When agents deviate to form coalition $M$ in partition $\P'$, agents in $C_1$ can arrange in any way. Again using Lemma \ref{Lem_merger_split_arrival}, we know that if agents of $C_1$ split they obtain smaller shares which increases the share of $M$. Hence, the worst utility that $M$ receives is under $\P'$. Thus, we have equation \eqref{Eqn_condition}, i.e. 

%We define $\underline{\lambda}_M^{\P'}$ to denote the minimum value that coalition $M$ can derive when agents arrange in configuration $\P'$

}
\begin{equation*}
 \lambda_{M}^{\mathcal{P}'} = \ulam_M^{\P'}  > \sum_{C_i \in M} \lambda_{C_i}^ \mathcal{P}.
\end{equation*}
which is same as the condition required under RB-IA rule. 

 Hence, there exists a configuration/payoff vector such that each of the members in $M$ obtain strictly better and thus, such a partition is not stable. \Cmnt{Hence, each agent in $M$ obtains better utility by equation \eqref{Eqn_Example_fair_rule_M} and thus such a partition is not stable.} \eop

\Cmnt{\begin{lemma}
\textit{ If $n$ agents split into a 2-partition with servers $k$ and $N-k$   such that $k>(N-k)$, then  coalition with $k$ servers gets more than $(k/N)\Lambda$ at WE.}
\label{Lem_dominating_coalition}
\end{lemma}}

\textbf{Proof of Theorem \ref{Thm_GC}: } i) There can be no merger from  $\P_G$,  and we only need to check if an appropriate split can block the given configuration  $(\P_G, \Phi)$ where     $\Phi$ is any payoff. \Cmnt{We have the following cases:

\textit{Case 1: When the agents are symmetric} 

This case comes under Case 2.

\textit{Case 1: When there exist a subset $S \subset \mathcal{N}$ and $S$ do not contains agent 1 such that $\sum_{i \in \mathcal{N} \backslash S}N_i \leq \sum_{i \in S}N_i$}}

a) We first consider all payoff vectors $\Phi$ that satisfy $$\sum_{i=2}^n \phi_i <\Lambda \left (   1-\frac{N_1}{N} \right ). $$
Let $S :=\{2, 3, \cdots, n\}$  be the coalition  made of all agents except agent $1$ and we will prove that this  coalition (split) will block the configuration of the form stated above. From Theorem \ref{Thm_WE}.(iii) since  $\ulam_1 < N_1/N \Lambda$,  coalition $S$ satisfies the following:\Cmnt{ Coalition $S$  satisfies both equations \eqref{Eqn_fair}   and  \eqref{Eqn_condition_S}, because by  using Theorem \ref{Thm_WE} part iii),   $\ulam_1 < N_1/N \Lambda$ which in turn implies } 
$$
\lambda_S^{\P'} =  \ulam_S > \Lambda \left (  1-\frac{ N_1}{N} \right )    ,  \mbox{ where }   {\P'}  := \{S, \{1\} \}. 
$$.

 Hence, there exists a configuration/payoff vector such that each of the members in $S$ obtain strictly better and thus, $\P_G$ is not stable for such $\Phi$.

b) Next, we consider all payoff vectors that satisfy \begin{eqnarray}
\sum_{i=2}^n \phi_i \geq \Lambda \left (  1-\frac{N_1}{N}  \right ). \label{Eqn_condition1}
\end{eqnarray}

For  $(\P_G, \Phi)$ to be stable, the payoff  vector should   satisfy $\sum_{i \in C} \phi_i \geq \underline{\lambda}_C = \lambda^{\P'}_C$  (with $\P' := \{C, {\cal N}\backslash C\} $),  for all $C$ (i.e., negating \eqref{Eqn_condition_S}). From Theorem \ref{Thm_WE} part iii),  any $C$ with more than $N/2$ servers satisfies  equation \eqref{Eqn_condition_S}, as $\lambda^{\P_G} = \Lambda$. We use a subset of such coalitions to complete the proof. Since $N_1$ is the agent with  maximum number of servers,
 $S_k := S \backslash \{k\} \cup \{1\}$ has $N_{S_k}  > N/2$ for any $k \ge 2$, do not satisfies equation \eqref{Eqn_condition_S}  by Theorem \ref{Thm_WE} part (iii) and will block the given payoff vector $\Phi$ if the following equations are not satisfied simultaneously:
 %we can replace one agent in $S$ with agent $1$ and write $(n-1)$ equations which satisfy equation \eqref{Eqn_condition_S} of the following form:
\begin{eqnarray*}
\phi_1+\phi_2+ \phi_3 \cdots+ \phi_{n-1} & > & \frac{N_1+\sum_{i=2}^{n-1} N_i}{N}\Lambda, \\
\phi_1+\phi_2+ \phi_3 \cdots+ \phi_{n-2} + \phi_n & > & \frac{N_1+\sum_{i=2}^{n-2} N_i+N_n }{N}\Lambda, \\
\vdots & > & \vdots \\
\phi_1+\phi_3+\phi_4 \cdots+ \phi_{n} & > & \frac{N_1+\sum_{i=3}^n N_i}{N}\Lambda.
\end{eqnarray*}
Adding these $(n-1)$ equations, we obtain the following:

\vspace{-4mm}
{\small\begin{eqnarray*}
(n-1)\phi_1+(n-2) \sum_{i=2}^n \phi_i & \hspace{-4mm} > \hspace{-4mm} & \frac{(n-1)N_1+(n-2)\sum_{i=2}^n N_i}{N} \Lambda, \\
\phi_1 + (n-2)\Lambda & \hspace{-4mm} > \hspace{-4mm} & \Bigg(\frac{N_1+(n-2)N }{N} \Bigg)\Lambda, \\
&&\text{ since } \sum_{i=1}^n \phi_i = \Lambda,   \sum_{i=1}^n N_i = N .\end{eqnarray*}}
Thus  for payoff vector $\Phi$ to be not  blocked  we require  that, $
\phi_1    >  \frac{N_1}{N}\Lambda $ and thus, $\sum_{i=2}^n \phi_i < \left( 1-\frac{ N_1}{N} \right)\Lambda$ which is not possible since we are considering payoff vectors that satisfy \eqref{Eqn_condition1}.

%However, if the agents in $S$ deviate, it satisfy equation \eqref{Eqn_Condition_S} and we also require \eqref{Eqn_fair}. Using Lemma \ref{Lem_dominating_coalition}
%, we know that 
%\begin{equation*}
%\ulam_S >\frac{\sum_{i=2}^n N_i}{N}\Lambda \text{ and } 
%\ulam_1 < \frac{N_1}{N}\Lambda.
%\end{equation*}
%Thus, we have
%$$
%\ulam_S > 1-\frac{ N_1}{N}\Lambda
%$$
%This implies that agent in $S$ can deviate and obtain better and thus payoff vector under GC is blocked coalition $S$.
Thus, $(\P_G,\Phi)$ is not a  stable configuration with any payoff vector $\Phi$.

\Cmnt{\textit{Case 2: When there exist a subset $S \subset \mathcal{N}$ containing agent 1 such that $\sum_{i \in \mathcal{N} \backslash S}N_i \leq \sum_{i \in S}N_i$}

Here we consider the case when there exist no such subset $S$ if agent $1$ is not included.}

ii) \textit{When $N_1 > \sum_{i \in \mathcal{N}; i \neq 1}N_i$}

In such a case, all coalitions that satisfy equation \eqref{Eqn_condition_S} must include agent 1. Thus, $\Phi$ should satisfy:
$$
\phi_1 > \max_{C} \ulam_C \text{ for all } C \subset \mathcal{N} \text{ satisfying \eqref{Eqn_condition_S} and containing agent 1.}
$$
Since no other coalition satisfy equation \eqref{Eqn_condition_S}, any payoff vector that satisfies the above equation \Cmnt{the above payoff vector}ensures configuration $(\P_G,\Phi)$  to be stable. % allocate agent 1 the maximum utility possible under all such coalitions. This will ensure that no other agent can lure agent 1 and thus under such an (initial) allocation vector, GC is stable. \Cmnt{consider an example with 3 agents with 8,4 and 2 servers respectively. If agent with 8 and 4 servers are allocated their maximum utilities possible partition $\P = \{\{1\}, \{2,\cdots,n\}\}$. From Lemma \ref{Lem_dominating_coalition}
%, we know that agent 1 obtains more than its proportional share under $\P$, say $x$. Thus, if the allocation rule allocates agent 1 a smaller amount than $x$ then it can deviate to $\P$. }
 \eop

\textbf{Proof of Theorem \ref{Thm_two_partition}:}  Any 2-partition $\P = \{C_1,C_2\}$ cannot be blocked by mergers since merger lead to $\P_G$ and \eqref{Eqn_condition_M} is not satisfied. Next we look at splits. Say $C_1 \in \mathbb{C}^*$. Then it follows from the definition of $\mathbb{C}^*$ that there exists no coalition $C \subset C_1$ such that it satisfies \eqref{Eqn_condition_S}. For any split of $C_2$ into $S$ and $C_2 \backslash S$, from Theorem \ref{Thm_WE}.(ii) we know that 
$$
\lambda_{C_2}^{\P} > \lambda_{S}^{\P'} + \lambda_{C_2 \backslash S}^{\P'}
$$ \Cmnt{$C_2$ obtains strictly better arrival rates. }Thus, there exists a payoff vector $\Phi$ that allocates strictly better to each player in $C_2$ which implies that such a split is not feasible.
Hence, any partition with one of the coalitions belonging to $\mathbb{C}^*$ is a stable partition under RB-IA rule.

Next, when each coalition has $N/2$ servers, the per-server share of each coalition equals $\Lambda/N$ at WE. However, if any subset of players (from any of the coalitions) deviate then the per-server share obtained by this coalition is strictly smaller than $\Lambda/N$ (from Theorem \ref{Thm_WE}.(iii)) and hence such a partition is stable.\eop % using same arguments as given above for showing infeasibility of any splits of $C_2$, such a partition is also stable.  \eop

\Cmnt{i) Let $k, N-k  \ne k^*$,  and let $\P = \{S,  {\cal N}\backslash S\}$ be the corresponding partition with $|S| = k$. 
Any blocking coalition $C$ anticipates minimal utility (pessimistic rule) and this happens when the rest of the agents are all together as in partition $\P' = \{C, {\cal N}\backslash C\}$. 
Consider $C$ such that $|C| = k^*$.
Thus  for any $i \in C$ by \eqref{Eqn_kstar}, 
$$
\frac{N_i}{N_C} \lambda_C^\P  >   \frac{N_i}{N_S} \lambda_S^\P    \mbox{ if }  i \in S.  
$$
Similar inequality follows with $ {\cal N}\backslash S$ in the right hand side if $i \in  {\cal N}\backslash S$.  Thus  \eqref{Eqn_blocking_coalition} is satisfied for any  $i \in C$ and hence $\P$ is blocked by $C$.

ii) Any 2-partition $\P = \{C_1,C_2\}$ cannot be blocked by mergers since merger lead to GC and equation \eqref{Eqn_condition_M} is not satisfied. It can also be not blocked by split since equation \eqref{Eqn_condition_S} is not satisfied because of our hypothesis.}  

\textbf{Proof of Theorem \ref{Thm_R_rule_stability}: } 
i) Consider any configuration  $(\P_G, \Phi \}$  with  GC.
The proof of this part can be split into two cases:

\textit{Case 1: When $N_1 \leq \sum_{i \in S; i \neq 1}N_i \text{ for some } S \subset \mathcal{N}$}

Under RB-PA rule for the configuration to be stable, we need to ensure that the  following system of equations are satisfied simultaneously.
\begin{eqnarray}
\sum_{i \in C} \phi_i \geq \ulam_C \text{ for all } C \subset \mathcal{N} \text{ and, }
\sum_{i \in \mathcal{N}} \phi_i  = \Lambda.  \label{Eqn_conditions}
\end{eqnarray} 
%From Lemma \ref{Lem_dominating_coalition}, we know that equation  \eqref{Eqn_condition_S} is  satisfied for coalitions with servers more than $N/2$. Thus, the allocation vector under GC should trivially satisfy those inequalities. 
However, a subset of these equations itself admit no feasible solution (as proved in Theorem \ref{Thm_GC}). Thus, such a system of equations does not have a solution and  hence  $(\P_G, \Phi \}$ is unstable for any payoff vector $\Phi$. \Cmnt{GC is hence unstable with any $\Phi$.}%  a contradiction has been shown in such a case.

\textit{Case 2: When $N_1 > \sum_{i \in \mathcal{N}; i \neq 1}N_i$}

Once again we need to satisfy \eqref{Eqn_conditions} to prove that $(\P_G, \Phi)$ is stable. In particular those equations will also have to be satisfied  for subsets $S$ such that $|S| = n-1$, and 
$1 \in S$.
If there exists a payoff vector $\Phi$  that satisfies all such conditions,  consider one such $S$ and say $j \notin S$.  Then  from \eqref{Eqn_conditions}:
$$\phi_j = \Lambda - \sum_{i \in S} \phi_i  \leq  \Lambda -\ulam_S 
= \ulam_{\{j\} },$$
If $\phi_j < \ulam_{\{j\}}$ for some $j$ then configuration $(\P_G, \Phi)$  is blocked by $\{j\}$ under RB-PA rule. Otherwise if $\phi_j = \ulam_{\{j\}}$ for all $j \in \mathcal{N}$ then $\sum_{i \in \mathcal{N}} \phi_i = \sum_{i \in \mathcal{N}} \ulam_{\{j\}} < \Lambda$ and thus \eqref{Eqn_conditions} is not satisfied.
\Cmnt{
Under anticipation rule \eqref{Eqn_alternate_rule}, we need to ensure feasibility of the following system of equations:
\begin{equation*}
\sum_{i \in S} \phi_i^{GC} > \ulam_{S} \text{ for all } S: |S|=(n-1).
\end{equation*}
If the solution to the above system of equations exist, i.e., if there exists a payoff vector $\Phi$ that satisfies the above set of equations for all $S$ then it indicates the existence of some agent $j$ outside coalition $S$ such that
$
\phi_j < \ulam_j
$ and this agent can deviate to block the configuration under GC.
}
Hence $(\P_G, \Phi)$ is unstable for any payoff $\Phi$. 

{\bf Proof of part ii):} Since the condition required for a merger to be successful under RB-PA rule is same as under RB-IA rule, the result follows from Theorem \ref{Thm_duo_mono}.

{\bf Proof of parts iii) and iv):} When the  payoff vector is given by equation \eqref{Eqn_PSA}, the RB-PA and RB-IA rules %allocation rules \eqref{Eqn_Example_fair_rule_M} or \eqref{Eqn_Example_fair_rule_S} and \eqref{Eqn_alternate_rule} 
are equivalent to each other. %Also, no merger can happen once we reach some 2-partition and we only need to consider splits. 
Thus, the result follows from Theorem \ref{Thm_two_partition}. % We know any merger cannot happen once a 2-partition is formed. So we need o check the stability of 2-partitions under split only. Assuming $\Phi$ to be the allocation rule under any 2-partition $\P= \{C_1, C_2\}$. The equations of following form need to be satisfied for a 2-partition to be stable:
%\begin{equation*}
%\sum_{i \in C} \phi_i \geq \ulam_C \text{ for all } C \subset C_i \text{ for } i \in \{1,2\}
%\end{equation*}
%{\color{red}However, the number of such conditions can be reduced by restricting to only those sub-coalitions which have more than $N/2$ servers.}

{\bf Proof of part v):}  The first part of this result follows from parts iii) and iv) of this theorem.

\Cmnt{{\color{red} \textbf{If part:} Any 2-partition is stable under RB-IA rule under the following cases:

\textit{Case 1: When \eqref{Eqn_condition_S} is not satisfied:}

Under this case, RB-PA rule with proportional payoff vector is equivalent to the RB-IA rule and thus, no subset of players can deviate and obtain better.

\textit{Case 2: When \eqref{Eqn_condition_S_pt2} is not satisfied:}

It is easy to see that if \eqref{Eqn_condition_S_pt2} is not satisfied then \eqref{eq:blocking_PA} is also not satisfied since $\ulam_Q^{\P} \leq \lambda_Q^{\P'}$

\textbf{Only if part:} If a partition is not stable under RB-PA rule with $\Phi_p^{\P}$ then there exists some coalition which will satisfy \eqref{Eqn_condition_S}. Following the arguments of Case 2, one can observe that \eqref{Eqn_condition_S_pt2} is also not satisfied.}}

 Moreover because of the continuity of $\Phi$, we have the next result.
  \eop
  
  \textbf{Proof of Theorem \ref{Thm_approx}: } The proof of Theorem \ref{Thm_approx} is available in main text in this report on page 8. \Cmnt{In this theorem, we first show that the result is true under second order approximation and the approximation converges to the actual using maximum theorem. Then we show that the result under second order approximation matches that with actual blocking probability.
  
  	The WE is obtained by equating the reciprocal of blocking probabilities \eqref{Eqn_PB} under second order approximation  we require zeros of the following: 
	\begin{eqnarray}
	1+\frac{k}{\lambda_1}+ \frac{k(k-1)}{\lambda_1^2} =  1+\frac{N-k}{\lambda_2}+\frac{(N-k)(N-k-1)}{\lambda_2^2} ,\nonumber 
	\end{eqnarray}
	which after some simple calculations leads  to the following fixed point equation (of $\lambda_1 \in [0, \Lambda]$ and $\lambda_2 := \Lambda-\lambda_1$):
	\begin{equation}
	 {\frac{\lambda_1}{k}= \frac{\lambda_2}{N-k} \Bigg(\frac{ 1+\frac{k}{\lambda_1}-\frac{1}{\lambda_1} }{  1+\frac{N-k}{\lambda_2}-\frac{1}{\lambda_2} } \Bigg)} .
	\label{Eqn_fixed_point_eq}
	\end{equation}
	Let $\psi (k) :=  {\hat \lambda}_k  / k $ where ${\hat \lambda}_k$ is the fixed point of the above function and observe that $\Psi (\cdot)$ represents similar function, but considering exact blocking probability  \eqref{Eqn_PB}.  For further analysis, we relax $k$ to be a real value between $(N/2, N)$. %We immediately have the following result:
	%\begin{lemma}
%\textit{i)	 There exists  a  ${\bar \Lambda}$ such that, 
%the function 	 $\psi$ is increasing with $k$, for any $\Lambda \ge {\bar \Lambda}$. \\
%ii) For any  $\Lambda \ge {\bar \Lambda}$,  under second order approximation, the unique maximizer in \eqref{Eqn_kstar} is given by   $k^*  = \sum_{i=1}^{n-1} N_i $.\\
%iii) For all such  $\Lambda$, the  partitions $\P = \{C,  {\cal N}\backslash C \}$,  with $  N/2 \le N_{C} \le k^*$ are the only  2-partitions that are stable, under second order approximation.}
%\label{heavy_traffic}
	   %\end{lemma}

{\bf Accuracy of the approximation}   
%We now prove that the above result is also true without approximation using \textit{maximum theorem} \cite{maximum}. 
We will first show that both the fixed points (under first order approximation and actual blocking probability) converge towards each other and that there exists  a  ${\bar \Lambda}$ such that,  stable partition considering true blocking probability equals that derived with second order approximation.

%\subsubsection{Equivalence of First Order Heavy Traffic Erlang B approximation and the exact Erlang B}
 Consider   $y \in [\epsilon, 1-\epsilon]$ for some small $\epsilon >0$ and define the following function: 
	
	\vspace{-4mm}
	{\small\begin{equation*}
	g(y, \theta ) = \left \{  \begin{array}{llll}
	 \sqrt{\frac{1}{\theta} } \Bigg(\sum_{j=0}^{k-1} \frac{( y/\theta)^{j-k}}{j!}k! \\ -  \sum_{j=0}^{(N-k)-1} \frac{ [(1-y)/\theta]^{j-(N-k)}}{j!}(N-k)! \Bigg)^2,  \hspace{-2mm}&\mbox{if } \theta  > 0 \\
	 (y^{-1}k-(1-y)^{-1}(N-k))^2  &\mbox{if } \theta = 0. 
	 \end{array}  \right . 
	\end{equation*}}
	Observe from \eqref{Eqn_PB}  that   $y = \frac{\lambda_1}{\Lambda}$ (the normalized WE)  is the unique zero of the  function  $g()$ when $\Lambda = 1/\theta$, uniqueness given by Theorem \ref{Thm_WE}, 
	and the solution by first order approximation is zero of $g$ when $\theta = 0$.  
	It is clear that $g$ is a jointly continuous mapping\footnote{If $\epsilon$ is such that the true WE does not fall in interval $[\epsilon, 1-\epsilon]$, 
 then we would have some other points as the minimizers of $g(.)$ (by continuity and compactness), but eventually (with large enough $\Lambda$) we will have unique zero of $g(.)$ which is derived from unique WE of the original problem.} over $[\epsilon, 1-\epsilon] \times [0, B]$ (for any $B < \infty$). Define,
	\begin{align*}
	g^{*}(\theta) & \triangleq  \max_{y \in [\epsilon,1-\epsilon]} g(y,\theta) \mbox{ and }
	y^{*}(\theta) & \triangleq \arg\max_{y \in [\epsilon,1-\epsilon]} g(y,\theta).
	\end{align*}
	Then, by Maximum Theorem, $y^{*}(1/\Lambda) \to y^{*}(0)$ as $\Lambda \to \infty$. In other words we have:
	\begin{equation}
 \frac{\lambda_1 (\Lambda)}{\Lambda} \to \frac{k}{N} \mbox{, or equivalently, }  \frac{1}{\Lambda} \Bigg|\frac{\lambda_1^*}{k}-\frac{\Lambda}{N}\Bigg| \to 0
 \label{Eqn_max_proof}
	\end{equation}
	Using exactly similar logic, one can show that the WE using  second order approximation also converges towards that of the first order approximation, and hence that  the differences between the  WE obtained using second order approximation and that   obtained using true blocking probability \eqref{Eqn_PB}  converge towards each other. 
	
	 \textit{Under second orrder approximation:}} \eop
	 
	 {\bf Proof of Lemma \ref{heavy_traffic}:} i) The equation \eqref{Eqn_fixed_point_eq} can be re-written as: (by replacing $\lambda_1/k$ with $\psi$ and $\lambda_2 = \Lambda-k\psi$) \Cmnt{Replace $\lambda_1/k$ with $\psi$ in equation \eqref{Eqn_fixed_point_eq} we have,}
	\begin{equation}
	\psi = \frac{\Lambda-k\psi}{N-k} \Bigg(\frac{1+\frac{1}{\psi}-\frac{1}{k\psi}}{1
		+\frac{N-k}{\Lambda-k\psi}-\frac{1}{\Lambda-k\psi}}\Bigg).
	\end{equation} 
	 Simplifying it further, we have \Cmnt{We can rewrite the above equation as:}
	\begin{eqnarray*}
	\psi \Bigg(1
	+\frac{N-k-1}{\Lambda-k\psi} \Bigg) & = & \frac{\Lambda-k\psi}{N-k} \Bigg(1+\frac{1}{\psi}-\frac{1}{k\psi}\Bigg).
	\end{eqnarray*}
	Relaxing $k$ to a real number and then differentiating the above equation with respect to $k$, we have
	\begin{eqnarray}
	\frac{\partial}{dk} \Bigg[\psi \Bigg(1
	+\frac{N-k-1}{\Lambda-k\psi} \Bigg)\Bigg] 
	&= &  \frac{\partial \psi}{ \partial k}\Bigg(1
	+\frac{N-k-1}{\Lambda-k\psi} \Bigg)+ \psi \Bigg(\frac{\partial}{dk} \left \{ \frac{N-k-1}{\Lambda-k\psi} \right \} \Bigg), \nonumber \\
	&  = &  \frac{\partial \psi}{ \partial k}\Bigg(1
	+\frac{N-k-1}{\Lambda-k\psi} \Bigg)+ \psi\Bigg(\frac{-(\Lambda-k\psi)+(N-k-1)(\psi+k\frac{\partial \psi}{ \partial k})}{(\Lambda-k\psi)^2}\Bigg), \nonumber \\
	& = &  \frac{\partial \psi}{ \partial k}\Bigg(1
	+\frac{N-k-1}{\Lambda-k\psi} + \frac{k\psi(N-k-1)}{(\Lambda-k\psi)^2} \Bigg)+  \psi\Bigg(\frac{-(\Lambda-k\psi)+(N-k-1)\psi}{(\Lambda-k\psi)^2}\Bigg). 
	\end{eqnarray}
	
	\begin{eqnarray}
	&&\frac{\partial}{dk} \Bigg[\frac{\Lambda-k\psi}{N-k} \Bigg(1+\frac{1}{\psi}-\frac{1}{k\psi}\Bigg)\Bigg]  \nonumber \\
	& = &  \Bigg(1+\frac{1}{\psi}-\frac{1}{k\psi}\Bigg) \frac{\partial}{dk} \left \{ \frac{\Lambda-k\psi}{N-k} \right \} + \Bigg(\frac{\Lambda-k\psi}{N-k}  \Bigg)\frac{\partial}{dk} \left \{\Bigg(1+\frac{1}{\psi}-\frac{1}{k\psi}\Bigg), \right \} \nonumber \\
	& = & \Bigg(1+\frac{1}{\psi}-\frac{1}{k\psi}\Bigg)\Bigg( \frac{-(N-k)\big(\psi+k\frac{\partial \psi}{ \partial k}\big)+(\Lambda-k\psi)}{(N-k)^2}\Bigg) +  \Bigg(\frac{\Lambda-k\psi}{N-k}  \Bigg)\Bigg(-\frac{1}{\psi^2}\frac{\partial \psi}{\partial k}+ \frac{1}{k^2\psi^2}\Big(\psi+k\frac{\partial \psi}{ \partial k}\Big)  \Bigg), \nonumber \\
	& = &  \frac{\partial \psi}{ \partial k}\Bigg[-\frac{k}{N-k}\Bigg(1+\frac{1}{\psi}-\frac{1}{k\psi}\Bigg)-\Bigg(\frac{1}{\psi^2}-\frac{1}{k\psi^2}\Bigg)\Bigg(\frac{\Lambda-k\psi}{N-k}  \Bigg) \Bigg] +  \Bigg[-\frac{\psi}{N-k}\Bigg(1+\frac{1}{\psi}-\frac{1}{k\psi}\Bigg) + \frac{(\Lambda-k\psi)}{(N-k)^2}\Bigg(1+\frac{1}{\psi}-\frac{1}{k\psi}\Bigg) \nonumber \\
	&  & +\frac{1}{k^2\psi}\Bigg(\frac{\Lambda-k\psi}{N-k}  \Bigg)\Bigg]. 
	\end{eqnarray}
	
	Combining the derivatives of LHS and RHS, we have
	\begin{eqnarray}
	\frac{\partial \psi}{ \partial k}\Bigg[1
	+\frac{N-k-1}{\Lambda-k\psi} + \frac{k\psi(N-k-1)}{(\Lambda-k\psi)^2}+\frac{k}{N-k}\Bigg(1+\frac{1}{\psi}-\frac{1}{k\psi}\Bigg)+\Bigg(\frac{1}{\psi^2}-\frac{1}{k\psi^2}\Bigg)\Bigg(\frac{\Lambda-k\psi}{N-k}  \Bigg) \Bigg] = \nonumber \\
	\Bigg[-\frac{\psi}{N-k}\Bigg(1+\frac{1}{\psi}-\frac{1}{k\psi}\Bigg) + \frac{(\Lambda-k\psi)}{(N-k)^2}\Bigg(1+\frac{1}{\psi}-\frac{1}{k\psi}\Bigg)+\frac{1}{k^2\psi}\Bigg(\frac{\Lambda-k\psi}{N-k}  \Bigg)-\psi\Bigg(\frac{-(\Lambda-k\psi)+(N-k-1)\psi}{(\Lambda-k\psi)^2}\Bigg)\Bigg]. \nonumber 
	\end{eqnarray}
	Hence,
	\begin{equation}
	\frac{\partial \psi}{ \partial k} = \frac{-\frac{\psi}{N-k}\Big(1+\frac{1}{\psi}-\frac{1}{k\psi}\Big) + \frac{(\Lambda-k\psi)}{(N-k)^2}\Big(1+\frac{1}{\psi}-\frac{1}{k\psi}\Big)+\frac{1}{k^2\psi}\Big(\frac{\Lambda-k\psi}{N-k}  \Big)-\psi\Big(\frac{-(\Lambda-k\psi)+(N-k-1)\psi}{(\Lambda-k\psi)^2}\Big)}{1
	+\frac{N-k-1}{\Lambda-k\psi} + \frac{k\psi(N-k-1)}{(\Lambda-k\psi)^2}+\frac{k}{N-k}\Big(1+\frac{1}{\psi}-\frac{1}{k\psi}\Big)+\Big(\frac{1}{\psi^2}-\frac{1}{k\psi^2}\Big)\Big(\frac{\Lambda-k\psi}{N-k}  \Big)}
	\end{equation}
	The denominator is always strictly positive. When $\Lambda \to \infty$, denominator tends to $1+\frac{k}{N-k} > 0$. The numerator can be re-written as:
	\begin{eqnarray}
	&&-\frac{\psi}{N-k}\Bigg(1+\frac{1}{\psi}-\frac{1}{k\psi}\Bigg) + \frac{(\Lambda-k\psi)}{(N-k)^2}\Bigg(1+\frac{1}{\psi}-\frac{1}{k\psi}\Bigg)+\frac{1}{k^2\psi}\Bigg(\frac{\Lambda-k\psi}{N-k}  \Bigg)-\psi\Bigg(\frac{-(\Lambda-k\psi)+(N-k-1)\psi}{(\Lambda-k\psi)^2}\Bigg) \nonumber \\
	&=& \frac{1}{N-k}\Bigg(1+\frac{1}{\psi}-\frac{1}{k\psi}\Bigg) \Bigg[-\psi+\frac{\Lambda-k\psi}{N-k} \Bigg]+\frac{1}{k^2\psi}\Bigg(\frac{\Lambda-k\psi}{N-k}  \Bigg)+\psi\Bigg(\frac{\Lambda-(N-1)\psi}{(\Lambda-k\psi)^2}\Bigg) \nonumber \\
	&=& \frac{1}{(N-k)^2}\Bigg(1+\frac{1}{\psi}-\frac{1}{k\psi}\Bigg) (\Lambda-N\psi )+\frac{1}{k^2\psi}\Bigg(\frac{\Lambda-k\psi}{N-k}  \Bigg)+\psi\Bigg(\frac{\Lambda-(N-1)\psi}{(\Lambda-k\psi)^2}\Bigg) \nonumber
	\end{eqnarray}
	%From equation \eqref{eqn_upper_bound}, we know the last term is positive while the first term is negative. Also, the second last term is positive. Hence in denominator, we want to show that as $\Lambda \to \infty$, the first term goes to zero while atleast one of the remaining terms is positive. In this direction, we want to show that $\Lambda \to Nx$ which makes first term zero and the second term positive.
Using equation \eqref{Eqn_max_proof} we have
	\begin{equation*}
	\frac{1}{(N-k)^2}\Bigg(1+\frac{1}{\psi}-\frac{1}{k\psi}\Bigg) (\Lambda-N\psi ) =\frac{1}{(N-k)^2}\Bigg(1+\frac{1}{\psi}-\frac{1}{k\psi}\Bigg) N\Lambda\frac{(\frac{\Lambda}{N}-\psi )}{\Lambda} \ \to \ 0.
	\end{equation*}
	\Cmnt{\begin{equation*}
		x\Bigg(\frac{\Lambda-(n-1)x}{(\Lambda-kx)^2}\Bigg) = \frac{x\Lambda}{(\Lambda-kx)^2}\Bigg(\frac{\Lambda-nx+x}{\Lambda}\Bigg) = \frac{nx\Lambda}{(\Lambda-kx)^2}\Bigg(\frac{\frac{\Lambda}{n}-x+\frac{x}{n}}{\Lambda}\Bigg) = \frac{x^2}{(\Lambda-kx)^2} 
		\end{equation*}}
	\begin{equation*}
	\frac{1}{k^2\psi}\Bigg(\frac{\Lambda-k\psi}{N-k}  \Bigg) = \frac{1}{k^2\psi}\Bigg(\frac{\Lambda-N\psi+N\psi-k\psi}{N-k}  \Bigg) = \frac{1}{k^2\psi}\Bigg(\frac{\Lambda-N\psi}{N-k}  \Bigg)+\frac{1}{k^2\psi}\Bigg(\frac{N\psi-k\psi}{N-k}  \Bigg) \to \frac{1}{k^2} 
	\end{equation*}
	since
	\begin{equation*}
	\frac{1}{k^2\psi}\Bigg(\frac{\Lambda-N\psi}{N-k}  \Bigg) = \frac{\Lambda}{k^2\psi}\Bigg(\frac{\Lambda-N\psi}{(N-k)\Lambda}  \Bigg) = \frac{N\Lambda}{k^2\psi}\Bigg(\frac{\frac{\Lambda}{N}-\psi}{(N-k)\Lambda}  \Bigg) \to 0.
	\end{equation*}
	From the similar arguments as above, we can observe that
	\begin{equation*}
	\psi\Bigg(\frac{\Lambda-(N-1)\psi}{(\Lambda-k\psi)^2}\Bigg) = \psi\Bigg(\frac{\Lambda-N\psi}{(\Lambda-k\psi)^2}\Bigg) + \psi\Bigg(\frac{\psi}{(\Lambda-k\psi)^2}\Bigg) \to \Bigg(\frac{\psi}{\Lambda-k\psi}\Bigg)^2 > 0
	\end{equation*}
	
	Hence, we have
	\begin{equation*}
	\frac{\partial \psi}{\partial k} > 0
	\end{equation*}
	 Thus, there exists a $\Lambda$ such that for all $\Lambda \geq \bar{\Lambda}$, $\psi$ is increasing with $k$.
	
	  ii) Since, $\psi$ is increasing with $k$ for any $\Lambda \geq \bar{\Lambda}$, the unique maximiser of \eqref{Eqn_kstar} is obtained at maximum possible of $k$ which is given by $\sum_{i=1}^{N-1} N_i$ (since $N_i$ are arranged in decreasing order). 
	
	 iii) When any coalition in $\P$, say $C_1$ with $N/2 \le N_{C_1} \leq k^*$ splits into $S$ and $C_1 \backslash S$ to form $\P'$, then using similar arguments as in Theorem \ref{Thm_WE}.(ii)
	$$
	\lambda_{C_1}^{\P} > \lambda_{S}^{\P'}+ \lambda_{C_1 \backslash S}^{\P'}.
	$$
	Thus, we can have a payoff vector $\Phi$ such that each player is strictly better in $C_1$. Hence, all such 2-partitions are stable under second order approximation.
	\Cmnt{\textit{Under true blocking probability:}
	
By the above arguments for any 2-partition  $\P = (C_1, C_2)$ with $N_{C_1}=k$, 
the 
 the WE obtained using second order approximation \eqref{Eqn_fixed_point_eq}, represented by   $ {\hat \lambda}_k$, and the one obtained  using the exact blocking probability  \eqref{Eqn_PB} converge towards each other for all $k$ as $\Lambda \to \infty.$
 Consider ${\bar \Lambda} $  further large in Lemma \ref{heavy_traffic} such that  (possible by finiteness)
$$\frac{1}{\Lambda}| {\hat \lambda}_k (\Lambda) - \lambda_k (\Lambda) | < \delta \mbox{ for all  possible } k, \mbox{ and for }   \Lambda \ge {\bar \Lambda}, $$ 
where $\delta >0$ is sufficiently small so that the conclusions of the theorem follow from that in Lemma \ref{heavy_traffic}  (possible by finitely many values of $k$), after observing that monotonicity of 
$\Psi$ with respect to $k$ for any given $\Lambda$ is equivalent to monotonicity of $\Psi / \Lambda$ with respect to $k$.} \eop

	\textbf{Proof of Lemma \ref{low_traffic}: } i) Recall the WE is obtained by equating  the reciprocal of the  blocking probabilities.  
	From \eqref{Eqn_PB},   at  low traffic   this can approximately be achieved by (solving for $\lambda_1$ which is a zero of the following):
	\begin{eqnarray}
	\frac{k!}{\lambda_1^{k}}  =  \frac{(N-k)!}{\lambda_2^{N-k}} \nonumber \mbox{ or }
%	\lambda_2^{N-k} & = & \frac{(N-k)!}{k!} \lambda_1^k \nonumber \\
	\lambda_2  =  \Bigg\{\frac{(N-k)!}{k!} \Bigg\}^{1/(N-k)} \lambda_1^{k/(N-k)}. \ 	\end{eqnarray}
	Using $\lambda_2 = \Lambda - \lambda_1$ we get 
	\begin{eqnarray}
	\frac{\lambda_1}{\Lambda}  =  \frac{1}{1+ \Big\{\frac{(N-k)!}{k!} \Big\}^{1/(N-k)} \lambda_1^{\frac{k}{N-k}-1}}.
	\end{eqnarray}
	Since $k> N-k$ we have $\frac{k}{N-k} >1$. As $\Lambda \to 0$, $\Big\{\frac{(N-k)!}{k!} \Big\}^{1/(N-k)} \lambda_1^{\frac{k}{N-k}-1} \to 0$ and thus, 
	\begin{equation}
	 \frac{\lambda_1}{\Lambda} \to 1 \mbox{ as } \Lambda \to 0.
	\label{Eqn_low}
	\end{equation}

\Cmnt{From this there exists ${\underline \Lambda} $  below which the given ${\underline k}^*$ optimizes \eqref{Eqn_kstar}. Observe this excludes all the partitions with $N_C = n/2$ exactly. Hence the second part is also through. }

  Suppose the coalition with higher number of servers has $\lfloor \frac{N}{2} \rfloor +m$ servers. From equation \eqref{Eqn_low} we have
	\begin{equation*}
	\frac{\lambda_1}{\Lambda \big( \lfloor \frac{N}{2} \rfloor + m \big)} \to \frac{1}{\big( \lfloor \frac{N}{2} \rfloor + m \big)} \mbox{ as } \Lambda \to 0.
	\end{equation*} Thus the arrival rate allotted to each server in this coalition as a fraction of total arrival rate lies in the interval
	$$
	\frac{1}{ \big( \lfloor \frac{N}{2} \rfloor + m \big)} - \epsilon < \frac{\lambda_1}{\Lambda k}< \frac{1}{\big( \lfloor \frac{N}{2} \rfloor + m \big)} + \epsilon.
	$$
	Choose an $\epsilon > 0$ such that the lower bound with minimum possible value of $m$, i.e., $\tilde{m}$ is greater than the upper bound for all $m > \tilde{m}$. Thus, we want to show:
	\begin{equation*}
	\frac{1}{\big( \lfloor \frac{N}{2} \rfloor + \tilde{m} \big)} - \epsilon > \frac{1}{\big( \lfloor \frac{N}{2} \rfloor + m \big)} + \epsilon  \quad \mbox{ for all } m > \tilde{m}.
	\end{equation*}
	It is sufficient to show that the upper bound for the least value of $m$ greater than $\tilde{m}$, i.e. $\hat{m}$ is smaller for some $\epsilon$, i.e.,
	\begin{eqnarray}
	\frac{1}{\big( \lfloor \frac{N}{2} \rfloor + \tilde{m} \big)} - \epsilon &> & \frac{1}{ \big( \lfloor \frac{N}{2} \rfloor + \hat{m} \big)} + \epsilon, \nonumber \\
	\frac{\hat{m}-\tilde{m}}{\big( \lfloor \frac{N}{2} \rfloor + \tilde{m} \big)\big( \lfloor \frac{N}{2} \rfloor + \hat{m} \big)} & > & 2\epsilon, \nonumber \\
	\frac{\hat{m}-\tilde{m}}{2\big( \lfloor \frac{N}{2} \rfloor + \tilde{m} \big)\big( \lfloor \frac{N}{2} \rfloor + \hat{m} \big)} & > & \epsilon \quad > \, 0. \nonumber
	\end{eqnarray}
	
	ii) The partitions with one of the coalitions having servers between $N/2 \le N_C \le k^*$ is stable can be proved using similar arguments as in Lemma \ref{heavy_traffic}.iii). For partition with $N_C > k^*$, players with $k^*$ servers can deviate together to obtain higher individual shares. This is because the players still obtain approximately same coalitional share but the number of players to share are less.
	\eop

	\Cmnt{\textbf{Under RB-PA when C has $N/2$ servers}
	when $C$ split into $S$ and $S'$ with their worths, then the blocking prob. is increased. Thus, at WE the worth of split coalition is smaller and hence it is not possible to allocate each of them better. also assume $\Phi_p$. then it can't be blocked as under proportional share both conditions are equivalent.}
	
	}{}

\Proofs{

\newpage

\onecolumn
	
	%\subsection{Solution Concept: Nash Equilibrium}
	%Consider any partition $\mathcal{P}$. If any player $i$ in coalition $C$ deviates to any other strategy profile, the utility of the deviated player is the minimum utility that it can obtain irrespective of the arrangement of players outside this coalition. We know that this player obtained higher utility in coalition and hence it is not better to deviate.
	%Thus, every partition is stable.
	
	%\subsection{Solution Concept: Coalitional Rationality}
	
%	We consider a partition to be coalitionally stable if no group of players can deviate together such that all of its members either gets higher utility or are indifferent among the two partitions, irrespective of the arrangement of players outside this coalition (when the other players are arranged as to hurt these players the most).

	%\noindent
	%{\color{blue} Here we can have two definitions, one when players are arranged so that they hurt you the most (which we are considering) and one when players are arranged so that they hurt us the least.}

	\Cmnt{

\newpage

\section{New Content}

\subsection{Results}
\begin{enumerate}
\item Partitions of size greater than 2 are blocked  under all conditions considered.
\item With condition $P$ and only merger or split allowed at once, we have one one history independent partition.
\item {\color{green}With condition $P$ and split and merge both allowed at the same time, once $C^*$ is formed, players can move from there as well even with fair shares for symmetric players.}
\item GC unstable for symmetric players under $P$ and $R$.
\item GC  unstable for $P$ even with asymmetric players when there exists a subset of players in $\cal{N}$ such that sum of their servers is either more or same as $N_1$. {\color{red} The other case left to prove.} If for $P$, then also for $R$.
\item For static Shapley Value, higher partitions, i.e. of size greater than $n/2$ are not stable.
\end{enumerate}	

\subsection{$.....$}
}	 }
	
	\Cmnt{
	\newpage

\section*{Can be used II}

\subsection{Coalition Formation Game}
%{\color{red} Define partition}.
Coalitional games involve a set of players $\mathcal{N} = \{1,2,\cdots,n\}$ who seek to form coalitions to improve their payoffs in the game. Usually, the worth of a coalition $C \subset \mathcal{N}$ is given by a characteristic function $\nu : C \to \mathbb{R}$. However, in games with externalities, the worth of a coalition depends not only on the actions of its members but also on the action of outside players, i.e., on the partition. A partition $\mathcal{P} = \{C_1,\cdots,C_m\}$ is the exhaustive and disjoint collection of subsets of $\mathcal{N}$ as given by equation \eqref{partition_define}.
 Hence in such a game, the worth of a coalition is given by a partition function $\nu : (C,\mathcal{P}) \to \mathbb{R}$ which associates a worth to each coalition $C \in \mathcal{P}$ corresponding to the partition  $\mathcal{P}$ it belongs to \cite{pessimistic}. For our queuing problem, $\nu(C, \mathcal{P}) = \lambda_{C}^\mathcal{P}$ is obtained using Wardrop Equilibrium.
 
 %To summarize, if a subset of providers say $C \subseteq \mathcal{N}$ agree to pool their resources then the blocking probability of this coalition of providers is given by:
 %\begin{equation}
 %B_{C}(N_C, a_{C,\mathcal{P}}) = \frac{\frac{{a_{C,\mathcal{P}}}^{N_C}}{N_C!}}{\sum_{j=0}^{N_C} \frac{{a_{C,\mathcal{P}}}^{j}}{j!}} \text{ where }
 %\end{equation}
 %where
  %$a_{C,\mathcal{P}}$ is the offered load to coalition $C$ in partition $\mathcal{P}$, and
 
 %\hspace{4mm} 
 %\noindent
 % $N_C$ is the number of servers with coalition $C$.
 
 %The arrival rate to all the coalitions in a partition is obtained by equating the blocking probabilities of each coalition in the partition according to the first principle of WE.

%
Under this context, 
consider a coalition planning to deviate from this partition. In order to obtain the value of this deviating coalition, one needs to anticipate the behaviour of outside players. %one needs to reconsider the definition of ``characteristic function". The characteristic function here gives the set of all possible allocations that 
%
%This is because when a coalition plans to block the given partition, its payoff is affected by the actions of players outside this coalition. Hence, the members of this coalition need to anticipate the behaviour of outside players.
%
The players can anticipate various arrangement of outside players as in \cite{pessimistic}. Some of them are:

\subsubsection*{Projection Rule}
Under this rule, the players in deviating coalition $C$ assume that the outside players do not change their arrangement when the players in $C$ deviate to improve their payoffs, i.e.,
\begin{equation}
\nu_p(C) = \nu(C,\mathcal{P}')
\end{equation}
where $\mathcal{P}' = \{C,C_1 \backslash C,C_2 \backslash C, \cdots, C_k \backslash   C\}$.

 The corresponding solution concept is the \textit{strong Cournot-Nash equilibrium} \cite{alpha-core}.

 As already mentioned, the solution concept depends on the allocation rule and we define it in the the following.
  An allocation rule is defined as follows:
\begin{equation}
\{\{\phi_{i,C}^{ \mathcal{P}}\}_{i \in C} ; \, \forall \, C \in \mathcal{P} \text{ and } \, \forall \, \mathcal{P}\} 
\end{equation}
where $C$ is any coalition in partition $\mathcal{P}$. In other words, an allocation rule divides the value of a coalition $C$ in partition $\mathcal{P}$ among its members such that 
\begin{equation}
\sum_{i \in C} \phi_{i,C}^\mathcal{P} = \nu(C,\mathcal{P}) ; \, \forall \, C \in \mathcal{P} , \, \forall \, \mathcal{P}
\end{equation}

Next, we define the solution concept corresponding to \textit{projection rule}.
\subsubsection*{Strong Cournot-Nash equilibrium}
Under this equilibrium, the deviating coalition

Alternatively, the players in deviating coalition may try to safeguard themselves from any irrational behaviour of outside players and follow the \textit{pessimistic rule}. 
\subsubsection*{Pessimistic Rule}
Under this rule, the players in deviating coalition $C$
 assume that the outside players arrange themselves to hurt $C$ the most, i.e., 
 \begin{equation}
 \nu_b(C) =  \min_{\mathcal{P}: C \in \mathcal{P}} \nu(C,\mathcal{P} )  % \text{ where }
 \end{equation} 
 where 
 $\nu(C,\mathcal{P} ) $ is the value of coalition $C$ under partition $\mathcal{P}$. 
 %This is commonly known as the \textit{pessimistic rule} \cite{pessimistic}.
  %
  The corresponding solution concept is \textit{$\alpha$-core} \cite{alpha-core}.
  
  Before defining \textit{$\alpha$-core}, we provide some important definitions.

\subsubsection*{$\alpha$-blocking}
An allocation $\{\phi_i^\mathcal{P}\}_{ i \in \mathcal{N}}$ is said to be $\alpha$ blocked by coalition $C$ if there exists another allocation $\{\phi_i^{'\mathcal{P}}\}_{ i \in \mathcal{N}}$
such that
\begin{equation}
\phi_j^{'\mathcal{P}} > \phi_j^{\mathcal{P}}  \, \forall \, j \in C \subset \mathcal{N}
\end{equation}
%and the above inequality is strict  for atleast one $j \in C$
 irrespective of the arrangement of players in $\mathcal{N}/C$.

\subsubsection*{$\alpha$-core}
A vector of feasible allocations which is not $\alpha$-blocked by any coalition is said to belong to $\alpha$-core.

\subsubsection*{Stable partition}
A partition is stable if the vector of feasible allocations under a given allocation rule  belong to $\alpha$-core.

%A coalition $C$ is said to \textit{$\alpha$-block} other coalitions if it  is strictly preferred by every agent $i$ in $C$ irrespective of the arrangement of players outside coalition $C$.
%A partition belongs to the \textit{$\alpha$-core} if no coalition can \textit{$\alpha$-block} it.

In this paper, we work with pessimistic rule.

\section*{Can Be Used III}

%An allocation rule is defined as follows:
%\begin{equation}
%\{\{\phi_{i,C}^{ \mathcal{P}}\}_{i \in C} ; \, \forall \, C \in \mathcal{P} \text{ and } \, \forall \, \mathcal{P}\} 
%\end{equation}
%where $C$ is any coalition in partition $\mathcal{P}$.

%The division of coalitional worth among its members is done using \textit{fair sharing scheme}.  

%We know by definition:
%\begin{equation}
%\sum_{i \in C} \phi_{i,C}^{\mathcal{P}} = {\lambda}_C^{ \mathcal{P}}
%\end{equation}
We say an allocation rule is \textit{fair} if it satisfies the following: ``Whenever two or more than two coalitions come together to form a single coalition, henceforth referred to as \textit{merger} obtains higher value, the division among the members of this merged coalition should become strictly better than before, for all the members of the merger." To be more precise,
if there exists a pairwise disjoint collection of coalitions $C_1,C_2, \cdots,C_k$ in any partition  $\mathcal{P}$  with merged coalition  $C := \cup_{i=1}^m C_i \text{ with } m \leq k$ having strictly better utility ``even when the players outside this coalition arrange themselves in any configuration $\tilde{\P}$", 
\begin{equation}
\sum_{i \in C} \phi_{i}^{\tilde{\P}}  > \sum_{i \in C} \phi_{i}^{\P} 
\end{equation}
where $\mathcal{P}'$ is the partition containing $C$ and the remaining players arranged to hurt coalition $C$ the most.

\noindent

\subsection{When is grand coalition unstable? }Following is the additional condition allocaton rule: If GC splits and one of the coalitions generate strictly better utility then each of the player sin coalition should get better. If this is satisfied then GC is not stable.

{\color{red} Corollary that proportional allocation satisfies this.}

%
%Hence, our aim is to study which partitions are stable when the service providers seek to form coalitions in order to maximize their individual arrival rates.

 %Hence, our aim is to find stable partitions such that the feasible allocations lie in \textit{$\alpha$-core}.
 
 \begin{lemma}
	When all except one coalition in partition merges, the new merger coalition obtanins higher utility
\end{lemma}

\noindent
\textbf{Proof: } Consider a partition $\mathcal{P} = \{C_1,C_2,\cdots, C_n\}$. At WE each coalition has equal blocking probability and hence when any two coalitions merge their blocking probability is reduced (compared to all other coalitions). Thus, all other coalitions lose some of their arrival rate. Now if the new merged coalition merges with some other coalition, its blocking probability is reduced and hence again other coalitions lose some of its utility to this coalition. Thus, continuing this consider a partition with all except $C_1$ to merge. Now the utility of merged coalition is the worst one when players in $C_1$ can arrange in any way. However, if $C_1$ splits, the players increase their blocking probability which is good for the merged coalition. Thus, we have that such a merger obtains higher sum utility. \eop

\section{Can be used III}
	\begin{lemma}
		\textit{Partitions with cardinality 2 are the only candidates for coalitionally stable partitions. }
		\label{coalitional_stable_partitions}
	\end{lemma}
	\noindent
	\textbf{Proof: } Here we show that partitions of either cardinality 1 or greater than 2 are not the stable ones.
	\begin{enumerate}
		\item \textbf{Partition of cardinality 1}
		
		In such a partition, each player gets $\frac{1}{n}$ share of total arrival rate. However, it is beneficial for all but one player to deviate together and get better.
		
		Thus, such a partition is not stable.
		
		\item \textbf{Partitions of cardinality greater than 2}
		
		Consider any partition of cardinality $k > 2$. Now at WE, each of the coalition has same blocking probability. If any number of coalitions (such that GC is not formed) come together, their blocking probability is reduced and hence, arrival rate is increased.
		
		Thus, such a partition can also be not stable.
	\end{enumerate}
	
	\Cmnt{\textbf{Proof:} Let total arrival rate be $\Lambda$ and each player be getting $\frac{\Lambda}{n}$. Now consider forming coalitions of size as in the lemma statement.
		The coalition with $k$ players can be seen as coalition of $(n-k)+(2k-n)$ players. The addition of $2k-n$ players reduces its blocking probability.
		
		\noindent
		Thus, the arrival rate of $k$-size coalition is more than $\frac{k}{n}$ share of total arrival rate.}
	
	\subsection{Heavy Traffic}
	The WE can be achieved by equating the reciprocal of blocking probabilities $R(N,a)$ which is given by:
	\begin{equation}
	R(N,a) = \sum_{j=0}^{N} \frac{N^{(j)}}{a^j}
	\label{reciprocal_ErlangB} 
	\end{equation}
	where
	$$
	N^{(j)} = N(N-1)\cdots(N-j+1), \quad j \geq 1.
	$$
	
	\noindent 
	From Lemma \ref{coalitional_stable_partitions} we know that the partitions with two colitions can only be the stable partitions. Also, the partition with same number of players in both coalitions cannot be the stable one. This follows from Lemma \ref{bigger_partition}.
	
	Hence, we consider a partition with two coalitions such that each coalition has $p$ and $n-p$ players respectively such that $p>n-p$. Each player has $\tilde{N}$ players such that
	$$
	\sum_{i=1}^{n} \tilde{N} = N
	$$
	Thus, the coalition with $p$ players has $p*\tilde{N} = k$ servers such that the other coalition has $N-k$ servers.
	\subsubsection{First Order Approximation}
	
	From equation \ref{reciprocal_ErlangB}, the first order heavy traffic appproximation for the WE  with $k$ servers in one coalition and is:
	\begin{eqnarray}
	1+\frac{k}{\lambda_1} & = & 1+\frac{N-k}{\lambda_2} \nonumber \\
	1+\frac{k}{\lambda_1} & = & 1+\frac{N-k}{\Lambda-\lambda_1} \nonumber \\
	\frac{k}{\lambda_1} & = & \frac{N-k}{\Lambda-\lambda_1} \nonumber \\
	k\Lambda & = & N\lambda_1 \nonumber %\\
	%\lambda_1 & = & \frac{k}{n} \Lambda \nonumber
	\end{eqnarray}
	Hence,
	\begin{equation}
	\boxed{\lambda_1  =  \frac{k}{N} \Lambda}
	\end{equation}
	\subsubsection{Second Order Approximation}
	From equation \ref{reciprocal_ErlangB}, the second order heavy traffic appproximation for the WE is:
	\begin{eqnarray}
	1+\frac{k}{\lambda_1}+ \frac{k(k-1)}{\lambda_1^2} & = & 1+\frac{N-k}{\lambda_2}+\frac{(N-k)(N-k-1)}{\lambda_2^2} \nonumber \\
	\frac{k}{\lambda_1}+\Bigg(\frac{k}{\lambda_1}\Bigg)^2- \frac{k}{\lambda_1^2} & = & \frac{N-k}{\lambda_2}+\Bigg(\frac{N-k}{\lambda_2}\Bigg)^2-\frac{(N-k)}{\lambda_2^2} \nonumber \\
	\frac{k}{\lambda_1} \Bigg( 1+\frac{k}{\lambda_1}-\frac{1}{\lambda_1} \Bigg) & = & \frac{N-k}{\lambda_2} \Bigg( 1+\frac{N-k}{\lambda_2}-\frac{1}{\lambda_2} \Bigg) \nonumber %\\
	%\frac{\lambda_1}{k} & = & \frac{\lambda_2}{n-k} \frac{\Big( 1+\frac{k}{\lambda_1}-\frac{1}{\lambda_1} \Big)}{ \Big( 1+\frac{n-k}{\lambda_2}-\frac{1}{\lambda_2} \Big)}
	%\label{fixed_point_eq}
	\end{eqnarray}
	Hence,
	\begin{equation}
	\boxed{\frac{\lambda_1}{k}= \frac{\lambda_2}{N-k} \Bigg(\frac{ 1+\frac{k}{\lambda_1}-\frac{1}{\lambda_1} }{  1+\frac{N-k}{\lambda_2}-\frac{1}{\lambda_2} } \Bigg)} 
	\label{fixed_point_eq}
	\end{equation}
	
	\subsubsection{Upper Bounds}
	Since $k > N-k$, from Lemma \ref{bigger_partition} we know $\frac{\lambda_1}{k} > \frac{\lambda_2}{N-k}$.
	From equation \eqref{fixed_point_eq}, we have
	
	\begin{eqnarray}
	\Bigg(\frac{ 1+\frac{k}{\lambda_1}-\frac{1}{\lambda_1} }{  1+\frac{N-k}{\lambda_2}-\frac{1}{\lambda_2} } \Bigg) & > & 1 \nonumber \\
	\Bigg( 1+\frac{k}{\lambda_1}-\frac{1}{\lambda_1} \Bigg)  & > & \Bigg( 1+\frac{N-k}{\lambda_2}-\frac{1}{\lambda_2} \Bigg) \nonumber \\
	\frac{k-1}{\lambda_1}  & > & \frac{N-k-1}{\Lambda-\lambda_1} \nonumber \\
	(k-1)(\Lambda-\lambda_1) & > & (N-k-1)\lambda_1 \nonumber \\
	k\Lambda-k\lambda_1-\Lambda+\lambda_1 & > & N\lambda_1-k\lambda_1-\lambda_1 \nonumber \\
	(k-1)\Lambda & > & (N-2)\lambda_1 \nonumber \\
	\lambda_1 &<& \frac{(k-1)\Lambda}{(N-2)} \nonumber \\
	\frac{\lambda_1}{k} & < & \frac{(k-1)\Lambda}{k(N-2)} \nonumber \\
	\frac{\lambda_1}{k} & < & \frac{\Lambda}{N-1} \nonumber
	\end{eqnarray} 
	The last inequality follows since the bound is increasing with $k$ and the maximum value of $k$ is $N-1$. Thus, we have:
	\begin{equation}
	\boxed{\frac{\Lambda}{N} < \frac{\lambda_1}{k} < \frac{\Lambda}{N-1}}
	\label{eqn_upper_bound}
	\end{equation}

	\subsubsection{Equivalence of First Order Heavy Traffic Erlang B approximation and the exact Erlang B}
	Define $y := \frac{\lambda_1}{\Lambda}$ such that $y \in [\epsilon, 1-\epsilon]$. Then, the reciprocal of blocking probability is defined as:
	\begin{equation*}
	\frac{\sum_{j=0}^{N_i} \frac{a_i^{j}}{j!}}{\frac{a_i^{N_i}}{N_i!}} = \sum_{j=0}^{N_i} \frac{a_i^{j-N_i}}{j!}N_i!
	\end{equation*}
	Now define function $g(y,\Lambda) :=$
	\begin{equation*}
	 \sqrt{\Lambda} \Bigg(\sum_{j=0}^{k-1} \frac{(\Lambda y)^{j-k}}{j!}k!-\sum_{j=0}^{(N-k)-1} \frac{[\Lambda(1-y)]^{j-(N-k)}}{j!}(N-k)! \Bigg)^2
	\end{equation*}
	Also, when $\Lambda \to \infty$, we have the following:
	\begin{equation*}
	g(y,\infty) =(y^{-1}k-(1-y)^{-1}(N-k))^2
	\end{equation*}
	Now, $g$ is a joint continuous mapping over $[\epsilon, 1-\epsilon] \times [0,\bar{\Lambda}]$ since it is the addition of two polynomial functions which are continuous. Hence,
	\begin{align*}
	g^{*}(\Lambda) & \triangleq  \max_{y \in [\epsilon,1-\epsilon]} g(y,\Lambda) \\
	y^{*}(\Lambda) & \triangleq \arg\max_{y \in [\epsilon,1-\epsilon]} g(y,\Lambda)
	\end{align*}
	Then, by Maximum Theorem, $y^{*}(\Lambda) \to y^{*}(\infty)$ as $\Lambda \to \infty$. Thus,
	\begin{equation}
	\boxed{\frac{\lambda_1^*(\Lambda)}{\Lambda} \to \frac{k}{N}}
	\end{equation}
	Hence, we have proved:
	\begin{equation}
	\boxed{\frac{1}{\Lambda} \Bigg|\frac{\lambda_1^*}{k}-\frac{\Lambda}{N}\Bigg| \to 0}
	\end{equation}
	{\color{red}We need to show that zero of the WE lies in the interval $[\epsilon,1-\epsilon]$}
	
	\subsubsection{Equivalence of Second Order Heavy Traffic Erlang B approximation and the exact Erlang B}
	To show:
	$$
	\frac{k(k-1)}{\lambda_1^2} - \frac{(N-k)(N-k-1)}{\lambda_2^2} \to 0 \text{ as } \Lambda \to \infty
	$$
	\noindent
	Define $y := \frac{\lambda_1}{\Lambda}$.
	\begin{eqnarray}
	\frac{k(k-1)}{\lambda_1^2} - \frac{(N-k)(N-k-1)}{\lambda_2^2} & = &\frac{k(k-1)}{(\Lambda y)^2} - \frac{(N-k)(N-k-1)}{(\Lambda(1-y))^2} \nonumber \\
	&= & \frac{1}{\Lambda^2} \Bigg[\frac{k(k-1)}{y^2} - \frac{(N-k)(N-k-1)}{(1-y)^2} \Bigg]
	\end{eqnarray}
	We know {\color{red}$ y \in [\epsilon, 1-\epsilon]$}.
	
	\subsection{Low Traffic}
	From equation \ref{reciprocal_ErlangB}, the low traffic approximation of reciprocal of Erlang B is given by:
	\begin{eqnarray}
	\frac{k!}{\lambda_1^{k}} & = & \frac{(N-k)!}{\lambda_2^{N-k}} \nonumber \\
	\lambda_2^{N-k} & = & \frac{(N-k)!}{k!} \lambda_1^k \nonumber \\
	\lambda_2 &=&  \Bigg\{\frac{(N-k)!}{k!} \Bigg\}^{1/(N-k)} \lambda_1^{k/(N-k)} \nonumber \\
	\Lambda-\lambda_1 &=&  \Bigg\{\frac{(N-k)!}{k!} \Bigg\}^{1/(N-k)} \lambda_1^{k/(N-k)} \nonumber \\
	\Lambda &=& \lambda_1 \Bigg[1+ \Bigg\{\frac{(N-k)!}{k!} \Bigg\}^{1/(N-k)} \lambda_1^{(2k-N)/(N-k)} \Bigg] \nonumber \\
	\frac{\lambda_1}{\Lambda} & = & \frac{1}{1+ \Big\{\frac{(N-k)!}{k!} \Big\}^{1/(N-k)} \lambda_1^{\frac{k}{N-k}-1}}
	\end{eqnarray}
	Since $k> N-k$ we have $\frac{k}{N-k} >1$. As $\Lambda \to 0$, $\Big\{\frac{(N-k)!}{k!} \Big\}^{1/(N-k)} \lambda_1^{\frac{k}{N-k}-1} \to 0$ and thus, 
	\begin{equation}
	\boxed{ \frac{\lambda_1}{\Lambda} \to 1 \mbox{ as } \Lambda \to 0}
	\label{low}
	\end{equation}
	
	\begin{lemma}
		\textit{There exists a $\bar{\lambda}_1(\epsilon)$  such that for all $\lambda_1 \leq \bar{\lambda}_1(\epsilon)$, partition with two coalitions with one of the coalitions having $\lfloor \frac{n}{2} \rfloor +1$ players is the coalitionally stable partition.}
	\end{lemma}
	
	\textbf{Proof: }	Suppose the larger coalition has $\lfloor \frac{n}{2} \rfloor +m$ players. From equation \eqref{Eqn_low} we have
	\begin{equation*}
	\frac{\lambda_1}{\Lambda \big( \lfloor \frac{n}{2} \rfloor + m \big)} \to \frac{1}{\big( \lfloor \frac{n}{2} \rfloor + m \big)} \mbox{ as } \Lambda \to 0
	\end{equation*} Thus the arrival rate allotted to each server in this coalition as a fraction of total arrival rate lies in the interval
	$$
	\frac{1}{ \big( \lfloor \frac{n}{2} \rfloor + m \big)} - \epsilon < \frac{\lambda_1}{\Lambda k}< \frac{1}{\big( \lfloor \frac{n}{2} \rfloor + m \big)} + \epsilon
	$$
	Choose an $\epsilon > 0$ such that the lower bound with $m=1$ is greater than the upper bound for akk $m \geq 2$. Thus, we want to show:
	\begin{equation*}
	\frac{1}{\big( \lfloor \frac{n}{2} \rfloor + 1 \big)} - \epsilon > \frac{1}{\big( \lfloor \frac{n}{2} \rfloor + m \big)} + \epsilon  \quad \mbox{ for all } m \geq 2
	\end{equation*}
	It is sufficient to show that the upper bound for $m=2$ is smaller for some $\epsilon$, i.e.,
	\begin{eqnarray}
	\frac{1}{\big( \lfloor \frac{n}{2} \rfloor + 1 \big)} - \epsilon &> & \frac{1}{ \big( \lfloor \frac{n}{2} \rfloor + 2 \big)} + \epsilon \nonumber \\
	\frac{1}{\big( \lfloor \frac{n}{2} \rfloor + 1 \big)\big( \lfloor \frac{n}{2} \rfloor + 2 \big)} & > & 2\epsilon \nonumber \\
	\frac{1}{2\big( \lfloor \frac{n}{2} \rfloor + 1 \big)\big(\lfloor \frac{n}{2} \rfloor + 2 \big)} & > & \epsilon \quad > \, 0 \nonumber
	\end{eqnarray}
	\eop

	%{\color{red} Check for this}

	\section{Asymmetric Players}
	
	We consider a system with $n$ independent service providers each with $N_i$ servers such that 
	$$
	\sum_{i=1}^{n} N_i = N
	$$
	
	\begin{lemma}
		If $n$ players split into a partition of two coalition with servers $k$ and $N-k$ such that $k>(N-k)$, then coalition with $k$ servers gets more than $\frac{k}{N}$ share of total arrival rate at WE.
		\label{share_asym}
	\end{lemma}
	
	\subsection{Division of Coalitional Worth: Proportional Allocation}
	
	\subsubsection{Solution Concept: Wardrop Equilibrium}
	
	\begin{lemma}
		{\it Partitions with cardinality 2 are the only candidates for coalitionally stable partitions.} 
	\end{lemma}
	\noindent
	\textbf{Proof: } Here we show that partitions of either cardinality 1 or greater than 2 are not the stable ones.
	\begin{enumerate}
		\item \textbf{Partition of cardinality 1}
		
		In such a partition, each server gets $\frac{1}{N}$ share of total arrival rate. However, it is beneficial for all but one player to deviate together and get better. (From Lemma \ref{share_asym})
		
		Thus, such a partition is not stable.
		
		\item \textbf{Partitions of cardinality greater than 2}
		
		Consider any partition of cardinality greater than 2. Now at WE, each of the coalition has same blocking probability. If any number of coalitions (such that GC is not formed) come together, their blocking probability is reduced and hence, arrival rate is increased.
		
		Thus, such a partition can also be not stable.
	\end{enumerate}
	
	\subsubsection*{Heavy Traffic}
	\noindent
	$\implies$ We also have that $\lambda_1/ k$ increases with $k$ where $k$ denotes number of servers in a coalition and $\lambda_1/k$ is the arrival rate per server.
	
	\noindent
	$\implies$ Since $\lambda_1/k$, i.e., arrival rate per server increases with $k$, i.e., number of servers in this coalition. We have the service provider with least number of servers is left out.
	
	\subsubsection*{Low Traffic}
	
	$\implies$ Equation \eqref{low} is still valid.
	
	\begin{lemma}
		\textit{ There exists a $\bar{\lambda}_1(\epsilon)$  such that for all $\lambda_1 \leq \bar{\lambda}_1(\epsilon)$, partition with two coalitions with one of the coalitions having $\lfloor \frac{N}{2} \rfloor +m$ servers  is the coalitionally stable partition, with $m$ being the least number of servers required so that  $\lfloor \frac{N}{2} \rfloor +m > N-\big(\lfloor \frac{N}{2} \rfloor +m \big)$.}
	\end{lemma}
	
	\subsubsection{Solution Concept: Nash Equilibrium}
	Consider any partition $\mathcal{P}$. If any player $i$ in coalition $C$ deviates to any other strategy profile, the utility of the deviated player is the minimum utility that it can obtain irrespective of the arrangement of players outside this coalition. We know that this player obtained higher utility in coalition and hence it is not better to deviate.
	Thus, every partition is stable.

\section{Can be used ....}

Some  allocation rules that satisfy the fair criteria are:

i) \textit{Proportional share based allocation rule (PSA) $\Phi_p$:} Under this rule,  the share of  value of any coalition (in any partition)  to any of its members is   proportional to value of its resources divided by the total value of the resources of all the members of the coalition. For example, in our queueing model:
\begin{eqnarray}
%\label{Eqn_PSA}
\Phi^{\P}_{p, i}  =  \frac{w_i(N_i)}{ \sum_{j \in C} w_i(N_j ) }   \lambda_C^\P \mbox{ for any } i  \in C \in \P {\color{red}\text{ and for all  } \P}.
\end{eqnarray}
These are fair as long $w_1(N_1) \ge w_2 (N_2) \cdots \ge w_n (N_n)$, where we assumed without loss of generality that $N_1 \ge N_2 \ge \cdots N_n.$ 

{\color{green}ii)  \textit{Extra value based allocation rule $\Phi_e$:}  A player in any merger  is allocated the value as it was obtaining before the merger plus some share of the extra value (positive/negative) derived by  the merger.  To define this rule precisely, we begin with single element partition $\P_a := \{ \{1\}, \cdots, \{n\}\}$, for which 
$$
\Phi_{e, i}^{\P_a}  := \lambda_{\{i\}}^{\P_a}  \mbox{ for each }  i.
$$
Say allocation rule  is defined for a partition $\P$ and say $\P' $ is obtained by merger $M$ of some of  the elements of $\P$ , then  the allocations for $\P'$ are given by: 
\begin{eqnarray}
\Phi^{\P'}_{e, i}  &=&  \Phi^\P_i +  \frac{1}{|M|}   \left ( \lambda_M^{\P'}  -   \sum_{C \subset M, C \in \P} \lambda_C^\P  \right )  \mbox{ for }  i \in M,  \nonumber \\
\Phi^{\P'}_{e, i} &=&  \Phi^\P_i +  \frac{  \lambda_C^{\P'}  -    \lambda_C^\P  }  {|C|},    \mbox{  for all  }  i \in  C \in  \P \cap \P'.
\end{eqnarray}
Say now $\P'$ is formed by splitting of an element   $M \in \P$ into $S$ and $S'$, then 
\begin{eqnarray}
\Phi^{\P'}_{e, i}  &=&  \Phi^\P_i +  \frac{1}{|M|}   \left ( \lambda_S^{\P'} +  \lambda_{S'}^{\P'}  -     \lambda_M^\P  \right )  \mbox{ for }  i \in  M,  \nonumber \\
\Phi^{\P'}_{e, i} &=&  \Phi^\P_i +  \frac{  \lambda_C^{\P'}  -    \lambda_C^\P  }  {|C|},    \mbox{  for all  }  i \in  C \in  \P \cap \P'.
\end{eqnarray}}

Next we present a result which eliminates the possibility of partitions of size more than 2, to be stable under any fair allocation rule.
%The following lemma eliminates the possibility of partitions of size more than 2 to lie in $\alpha$-core under fair sharing schemes.
\begin{theorem}
%\label{Thm_duo_mono}
	\textit{Partitions of cardinality greater than 2 are not stable under a fair allocation rule, for our queueing system.}

\end{theorem}
\noindent
\textbf{Proof: } See Appendix B. \eop

\section{Can be used ... }

\newpage

\subsection{UGC  rule {\color{red} probably write this subsectiton only for PSA?}}  Let the  grand coalition (GC) be denoted by  $\P_G = \{  {\cal N} \}$. We say  any  allocation rule is   Unfair GC (UGC) under $\Phi$ if there exists   a coalition $C$ such that the following is true {\color{red} with   $\P := \{C, {\cal N} \backslash C \}$ (under PSA), or for every partition  $\P$ containing $C$ for general allocation rule }   
 \begin{equation}
\ \Phi_i^\P  >   \Phi_i^{\P_G}      \mbox{ for each } i \in  C \subset \P.
  \label{Eqn_GC_fair}
\end{equation}
%and say  one of the coalitions generates strictly better utility {\color{red}under $\Phi$}, i.e., 
%say
%$$
%\sum_{i \in C }  \Phi_i^{\P_G}  <   \lambda_C^\P  \mbox{ for some } C \in \P.
%$$
 %then we say {\it the allocation rule is GC-fair},  if  each of the players of $C$     get strictly better:   
 %$$
  %\Phi_i^{\P_G}  <   \ \Phi_i^\P  \mbox{ for each } i \in  C \in \P,
%$$
%{\color{blue} Observe that the above rule depends on $\Phi$}
The following result is immediate  from the definition:
\begin{lemma}
\textit{GC is a stable partition if and only if the allocation rule is not UGC.
}  
\label{monopoly}
\end{lemma}
\noindent
\textbf{Proof: } The proof is straightforward. \eop 

We have already shown that PSA rule falls under fair allocation rule. Next we show that it is in fact, UGC as well. This result is an immediate consequence of the following lemma:

\begin{lemma}
\textit{i) If $n$ players split into a 2-partition with servers $k$ and $N-k$ where $N:= \sum_{i \in \mathcal{N}} N_i$ such that $k>(N-k)$ , then under PSA rule, coalition with $k$ servers gets more than $(k/N)\Lambda$ at WE.}

{\color{red} Requires extra condition }
\textit{ii) PSA rule is UGC.}
%\label{Lem_dominating_coalition}
\end{lemma}
\textbf{Proof:} See Appendix B. \eop

{\color{red} With 2 players having same number of servers, GC and all-alone, both partitions are stable.}

	\section{Proportional Fair Allocation rule  (PSA)}
	
It is now clear that only 2-partitions are stable for our queueing model under PSA. In this section we derive the partitions among  the 2-partitions  that are stable. 

Let $N := \sum_i N_i$ be the total number of servers of all the  players and assume  $N_1 \ge N_2 \ge \cdots N_n$, without loss of generality.  Any 2-partition $\P = \{C_1, C_2\}$ can be  identified uniquely by $k := |C_1| $ the number of servers  of its  first coalition.   For simpler notations,  we  let $\lambda_k := \lambda_{C}^\P$, when $|C| = k$,  and this definition is possible because all the servers are identical. 
Our aim now  is to find  the values of  $k$ that renders the 2-partition stable. 

Let $\Psi(k ; \Lambda)$ represent $\lambda_k / k$ for any given $\Lambda$. By Theorem \ref{Thm_WE},   the function $\Psi(k ; \Lambda)$  equals $1/k$ times the unique  zero  of the following function  of $\lambda$ (see \eqref{Eqn_WE_properties}):
$$
h(\lambda) :=   \frac { \lambda^k } { k! }   \sum_{j=0}^{N-k} \frac{ (\Lambda-\lambda)^j }{j!}   -  \frac{{(\Lambda-\lambda)}^{N-k}}{(N-k)!}     \sum_{j=0}^{k} \frac{\lambda^j} {j!}.
 $$
 Observe here that $\lambda_{N-k} = \Lambda - \lambda_k$. We now have the description of the stable partition under PSA.
 \begin{lemma}{\bf [Stable partition]}
The only  value of $k$  that renders the  partition $\P = \{S,  {\cal N}\backslash S\}$ (with $|S| = k$)   stable under PSA rule \eqref{Eqn_PSA}, is given  by:
\begin{eqnarray}
 k^* (\Lambda) := \arg \max_{k: k = |C|, C \subset {\cal N} }   \Psi(k ; \Lambda). %\label{Eqn_kstar_can_beused}
\end{eqnarray}
\end{lemma}
{\bf Proof:} 
Let $k, N-k  \ne k^*$,  and let $\P = \{S,  {\cal N}\backslash S\}$ be the corresponding partition with $|S| = k$. 
Any blocking coalition $C$ anticipates minimal utility (pessimistic rule) and this happens when the rest of the players are all together as in partition $\P' = \{C, {\cal N}\backslash C\}$. 
Consider $C$ such that $|C| = k^*$.
Thus  for any $i \in C$ by \eqref{Eqn_kstar}, 
$$
\frac{N_i}{N_C} \lambda_C^\P  >   \frac{N_i}{N_S} \lambda_S^\P    \mbox{ if }  i \in S.  
$$
Similar inequality follows with $ {\cal N}\backslash S$ in the right hand side if $i \in  {\cal N}\backslash S$.  Thus  \eqref{Eqn_blocking_coalition} is satisfied for any  $i \in C$ and hence $\P$ is blocked by $C$ under PSA.  \eop

Further analysis (of $k^*$) is carried out in  heavy  and light traffic regimes. For the rest of the cases we derive the results using numerical computations. 

{\color{red}Observe here that any reasonable scheduler (e.g., Shapely valued based etc) would be proportional fair schedulers  for symmetric case, i.e., when $N_i$ is the same for all $i$.  
}

\newpage

\section{Maximal Dynamics }

{\color{red} This part is not yet verified/proved completely !!

There can be multiple moves (merger/split) at any step from any given  partition $\P $ and allocation vector $\Phi^\P$ which can be successful. 
We obtain further analysis by considering that merger/split happens in some maximal sense, i.e.,  as below: 
\begin{eqnarray}
C^* (\P, \Phi)  \in \arg \max_{ C, satisfies \eqref{Eqn_condition_M} or \eqref{Eqn_condition_S} and \eqref{Eqn_fair}}     \frac{  \ulam_C  - \sum_{C_j \cap C \ne \emptyset }  \lambda_C \frac{N_{C_j \cap C}}{N_{C_j} }  } { N_C }  \label{Eqn_max_dynamics}
\end{eqnarray}
With this kind of maximal movement, we would show below that the dynamics settles down to 2-partitions, no matter where they started.  It may not settle, but might keep toggling majorly between some of the two-partitions, possibly till it hits upon $\P^*$.

%\hline

============

We make the following assumption.

{\bf Random Dynamics:} Let me consider random picking up of the possible better partitions.  That means, in every step, if a Partition is not stable then one of the  new partitions with merger/split that satisfy \eqref{Eqn_condition_M} or \eqref{Eqn_condition_S} and \eqref{Eqn_fair}  will be picked-up (equally likely or with a probability distribution that supports all possibilities). \\
{\bf Result:} Then the  random dynamics converges to one of the stable partitions of Theorem \ref{Thm_R_rule_stability}  (either $\{ C^*, {\cal N}-C^*\}$  or one of the following 2-partitions  $\{ C, {\cal N}-C\}$ such that $C \cap C^* \ne C^*$ for any $C^*.$)
with probability one in finite time.  \\
{\bf Proof :} This result I think is true even without {\bf A}.1 and is triivally true because of random dynamics.  Need to check..

{\bf Result 2:} Then the maximal  dynamics \eqref{Eqn_max_dynamics} either converges to  one of the stable partitions of Theorem \ref{Thm_R_rule_stability}  ($\{ C^*, {\cal N}-C^*\}$  or  to one of the following 2-partitions  $\{ C, {\cal N}-C\}$ such that $C \cap C^* \ne C^*$ for any $C^*$), or will toggle between 2 Partitions and 3 partitions.  \\
{\bf Proof:}  I still need to work on one part (i.e., under maximal dynamics does it come down to 2 P)?  Once it comes down I have proved that it either toggles or settles at one of the partitions. 
 
 Under A.1, May be that when we start with general partition, the maximal dynamics converge a toggling pattern in  the lower $n- |C^*|-1$ levels. 
 
============

 }

\twocolumn
}


\begin{thebibliography}{9}

%\bibitem{TR}
%Technical report downloadable from
%\textit{https://www.ieor.iitb.ac.in/files/faculty/kavitha/CoopQueue.pdf}.

	\bibitem{WE}
	Correa, Jos{\'e} R and Stier-Moses, Nicol{\'a}s E,
	\textit{Wardrop equilibria}
	Wiley encyclopedia of operations research and management science, 2010.
	
	\bibitem{aumann1961}
	Aumann, Robert J,
	\textit{The core of a cooperative game without side payments.}
	Transactions of the American Mathematical Society, vol. 98, no. 3, pp. 539--552, 1961.
	
	\bibitem{alpha-core}
	Martins-da-Rocha, Victor Filipe and Yannelis, Nicholas C,
	\textit{Non-emptiness of the alpha-core.}
	Funda{\c{c}}{\~a}o Getulio Vargas. Escola de P{\'o}s-gradua{\c{c}}{\~a}o em Economia, 2011.
	
	\bibitem{pessimistic}
	Bloch, Francis and Van den Nouweland, Anne,
	\textit{Expectation formation rules and the core of partition function games,}
	Games and Economic Behavior, vol. 88, pp. 339--353, 2014.
	
	\bibitem{Shiksha_Perf}
	Shiksha Singhal and Veeraruna Kavitha,
	\textit{Coalition Formation Resource Sharing Games in Networks}
	Accepted in Performance Evaluation, 2021.
	
	\bibitem{CFGs}
	Hajdukov{\'a}, Jana,
	\textit{Coalition formation games: A survey}
	International Game Theory Review, vol. 8, no. 04, pp. 613--641, 2006.
	
	\bibitem{saad2009}
	Saad, Walid and Han, Zhu and Debbah, M{\'e}rouane and Hjorungnes, Are and Basar, Tamer
	\textit{Coalitional game theory for communication networks}
	IEEE signal processing magazine, vol. 26, no. 5, pp. 77--97, 2009.
	
	\bibitem{karsten}
	Karsten, Frank and Slikker, Marco and Van Houtum, Geert-Jan
	\textit{Resource pooling and cost allocation among
independent service providers}
Operations Research, vol. 63, no. 2, pp. 476--488, 2015, INFORMS.

\bibitem{karsten_loss}
Karsten, Frank and Slikker, Marco and van Houtum, Geert-Jan
\textit{Analysis of resource pooling games via
a new extension of the Erlang loss function}
Tech. Rep., BETA working paper 344, Eindhoven University of Technology, 2011.

	\bibitem{maximum}
Sundaram, Rangarajan K and others,
\textit{A first course in optimization theory}
Cambridge university press, 1996.

\bibitem{formulas}
http://www.columbia.edu/~ww2040/CallCenterF04/questions1a.pdf
\end{thebibliography}
\end{document}